\documentclass{article}

\usepackage{PRIMEarxiv}

\usepackage{amsmath}
\usepackage{booktabs}
\PassOptionsToPackage{hyphens}{url}\usepackage[hidelinks]{hyperref}
\usepackage{algorithm}% http://ctan.org/pkg/algorithms
\usepackage{algpseudocode}% http://ctan.org/pkg/algorithmicx
\usepackage{amssymb}
\usepackage{comment}     
\usepackage{float}
\usepackage{hyperref}
\usepackage{lscape}
\usepackage{multirow}
\usepackage[flushleft]{threeparttable}
\usepackage[english]{babel}
\usepackage{graphicx}
\usepackage{amsthm}
\usepackage{siunitx} % Load the siunitx package for units

\newtheorem*{remark}{Remark}

\usepackage[table]{xcolor}
\usepackage{xcolor}
\definecolor{c1}{rgb}{0,0,1} % blue
\definecolor{c2}{rgb}{0,0.1,0.9} % light blue
\definecolor{c3}{rgb}{0.7,0,0} % dark red
\definecolor{c4}{rgb}{0,0,0} % black
\hypersetup{
    linkcolor={c4}, % internal links
    citecolor={c2}, % citations
    urlcolor={c3} % external links/urls
}
\title{An Optimization Framework to Personalize Passive Cardiac Mechanics
\thanks{\textit{\underline{Citation}}: 
\textbf{Shi et al. An Optimization Framework to Personalize Passive Cardiac Mechanics. Under Review }} 
}

\author{
  Lei Shi \\
  Department of Mechanical Engineering \\
  Columbia University \\
  New York\\
  \texttt{ls3374@columbia.edu} \\
   \And
  Hiroo Takayama \\
  Department of Cardiothoracic and Vascular Surgery \\
  Columbia University Irving Medical Center \\
  New York\\
  \And 
  Ian Chen \\
  Divison of Cardiovascular Medicine \\
  Stanford University \\
  Stanford\\
  \And
  Vijay Vedula \\
  Department of Mechanical Engineering \\
  Columbia University \\
  New York\\
  \texttt{vv2316@columbia.edu} \\
}

\begin{document}
\maketitle

\begin{abstract}
Personalized cardiac mechanics modeling is a powerful tool for understanding the biomechanics of cardiac function in health and disease and assisting in treatment planning. However, current models are limited to using medical images acquired at a single cardiac phase, often limiting their applicability for processing dynamic image acquisitions. This study introduces an inverse finite element analysis (iFEA) framework to estimate the passive mechanical properties of cardiac tissue using time-dependent medical image data. The iFEA framework relies on a novel nested optimization scheme, in which the outer iterations utilize a traditional optimization method to best approximate material parameters that fit image data, while the inner iterations employ an augmented Sellier’s algorithm to estimate the stress-free reference configuration. With a focus on characterizing the passive mechanical behavior, the framework employs structurally based anisotropic hyperelastic constitutive models and physiologically relevant boundary conditions to simulate myocardial mechanics. We use a stabilized variational multiscale formulation for solving the governing nonlinear elastodynamics equations, verified for cardiac mechanics applications. The framework is tested in myocardium models of biventricle and left atrium derived from cardiac phase-resolved computed tomographic (CT) images of a healthy subject and three patients with hypertrophic obstructive cardiomyopathy (HOCM). The impact of the choice of optimization methods and other numerical settings, including fiber direction parameters, mesh size, initial parameters for optimization, and perturbations to optimal material parameters, is assessed using a rigorous sensitivity analysis. The performance of the current iFEA is compared against an assumed power-law-based pressure-volume relation, typically used for single-phase image acquisition.
\end{abstract}

% keywords can be removed

%%%%%%%%%%%%%%%%%%%%%%%%%%%%%%%%%%%%%%%%%%%%%%%%%%%%%%%%%%%%%%%%%%%%%%%%%%%
%------------------------    INTRODUCTION   -------------------------------
%%%%%%%%%%%%%%%%%%%%%%%%%%%%%%%%%%%%%%%%%%%%%%%%%%%%%%%%%%%%%%%%%%%%%%%%%%%
\section{Introduction}\label{sec:intro}

Computer-based heart modeling stands out as a powerful, non-invasive tool for delving into the intricacies of cardiac function. These models play a crucial role in elucidating detailed structure-function relationships, proving indispensable for advancing basic research. Recent improvements in medical imaging and computational methods have extended the utility of these models, making them invaluable for understanding cardiac disease and dysfunction, device design, and treatment planning.\cite{khodaei2023enhancing,kadem2022hemodynamic,mittal2016computational} Simplified models have been used previously to develop constitutive models of the myocardium and to understand the role of mechanics and blood flow on cardiac function in both health and disease.\cite{arts1979model,arts1992description,guccione1991passive,guccione1993mechanics,usyk2000effect,costa2001modelling,holzapfel2009constitutive,vedula2015hemodynamics,vedula2016effect,seo2014effect,chnafa2015using} However, with rapid advances in medical imaging enabling high-quality image acquisition with superior spatial and temporal resolution and low acquisition times and radiation dosage, the emphasis is shifting to developing computational models tailored for an individual.\cite{crozier2016image,marx2022robust,balaban2017high,stimm2022personalization,arevalo2016arrhythmia} These patient-specific models must accurately mirror the distinctive characteristics of an individual's heart, ensuring decisions regarding personalized treatments are well-founded. However, aligning these computer models with clinical observations, commonly known as "model personalization" or "digital twinning," poses a formidable challenge.\cite{marx2022robust} The heart's functioning encompasses various strongly coupled physical processes, including electrophysiology, myocardial mechanics, and blood flow.\cite{quarteroni2017cardiovascular,verzicco2022electro} While accurate computer models exist for each process,\cite{vedula2015hemodynamics,finsberg2018efficient,gillette2022personalized} constructing a comprehensive model involves numerous parameters, some of which remain beyond direct measurement. Consequently, these parameters must be indirectly estimated from observable clinical data, introducing a layer of uncertainties.\cite{hirschvogel2017monolithic,salvador2023fast,fedele2023comprehensive}

Identifying myocardial material parameters holds paramount significance in diverse applications within cardiac biomechanics research. As the muscle responsible for the heart's pumping function through rhythmic contraction and relaxation,\cite{janssen2010myocardial} the myocardium exhibits complex structural and mechanical behavior that directly impacts its overall function.\cite{holzapfel2009constitutive} Personalizing the mechanical behavior of the myocardium using time-resolved image data proves highly valuable for analyzing myocardial pathology and evaluating treatment plans. It makes it feasible to simulate and predict how the myocardium deforms and interacts with its surrounding structures under various loading conditions, specific to the individual's pathology. Further, characterizing the material parameters at the baseline will allow for evaluating changes in myocardial deformation and stress profiles in response to therapeutic and surgical interventions. Ultimately, this information proves invaluable for simulation-guided disease diagnosis and prognosis and plan treatment with applications including myocardial infarction, hypertrophic cardiomyopathy, arrhythmia, heart failure, etc.

Myocardial biomechanics has been thoroughly characterized using structurally based constitutive models, parameterized to fit experimentally measured stress-strain profiles.\cite{guccione1991passive,guccione1993mechanics,holzapfel2009constitutive} Personalizing the mechanical behavior of the myocardium, therefore, involves obtaining patient-specific material parameters of a \textit{chosen} constitutive model by combining techniques from medical image processing and computational modeling. Previous works focused on developing methods for identifying myocardial material parameters using human or animal models, either \textit{in vivo} or \textit{ex vivo}. Augenstein et al. applied diffusion tensor magnetic resonance imaging (DT-MRI) combined with pressure recordings to estimate passive material properties in pig hearts.\cite{augenstein2005method} Gao et al. developed a multi-step nonlinear least squares optimization protocol to estimate the passive material parameters of a left ventricular myocardium from MRI-based volumes and strain data in healthy individuals.\cite{gao2015parameter} Finsberg et al. employed a gradient-based optimization method to obtain biventricular myocardial material parameters.\cite{finsberg2018efficient} Nair et al. conducted isolated finite element analysis (FEA) on the rabbit left ventricle (LV) and obtained passive material parameters through strain matching.\cite{nair2007optimizing} Xi et al. introduced a reduced-order unscented Kalman filter method to estimate passive material parameters by matching the FEA model to three-dimensional (3D) tagged MRI data.\cite{xi2011myocardial,xi2013estimation,xi2014understanding} Marx et al. developed an efficient technique to estimate passive material parameters in a human LV model by fitting their FEA model to the Klotz curve, approximating the ventricular end-diastolic pressure-volume relation (EDPVR).\cite{marx2022robust,klotz2006single} 

While these commendable efforts have allowed us to progress toward creating accurate digital twins, some either lacked or considered limited image information,\cite{marx2022robust,augenstein2005method,gao2015parameter,nair2007optimizing,xi2011myocardial,xi2013estimation,xi2014understanding} or focused on fitting a few material parameters.\cite{marx2022robust,finsberg2018efficient} Moreover, the primary focus of these investigations has been on ventricular mechanics, most studying LV alone. On the other hand, the left atrium (LA) remains relatively less studied despite its clinical significance in complex pathological conditions, such as cerebral stroke due to atrial fibrillation and flutter.\cite{augustin2020impact,aronis2019understanding,carpenito2021central} A notable gap remains due to a lack of a framework for personalizing the FEA model to match \textit{in vivo} image-based deformation acquired at several phases during the cardiac cycle. Developing such a framework should involve adjusting multiple material parameters to robustly align with the myocardial cavity volumes and local deformation extracted from the images.

Myocardial mechanics can be divided into a passive component and an active contraction component.\cite{guccione1993mechanics,goktepe2014generalized} Each component needs to be tuned separately to align with image-based deformation during the entire cardiac cycle. In this study, our focus centers on passive mechanics, aimed at developing an inverse FEA framework for estimating material properties of the passively deforming myocardium, along with obtaining the reference stress-free configuration simultaneously, using a patient's time-resolved 4D (3D + time) computed tomographic (CT) data. We apply this framework to characterize the biventricular and LA tissue’s passive mechanical behavior in a subject with a normal myocardium (control). We further demonstrate the framework's robustness by employing it to study the biventricular myocardium in three patients with hypertrophic obstructive cardiomyopathy (HOCM). HOCM patients manifest a thickened septum obstructing blood flow through the left ventricular outflow tract (LVOT).\cite{nishimura2017hypertrophic} Characterizing the myocardial material parameters in these patients may assist in devising personalized simulation-driven treatment planning approaches that relieve obstruction. While the current study is centered on characterizing passive mechanics, in a follow-up investigation, we will employ the estimated passive material parameters to tune the active contraction component using a coupled multiscale cardiac mechanics modeling framework.\cite{brown2024modular}

\begin{figure}[H]
\centering
\includegraphics[width=0.7\textwidth]{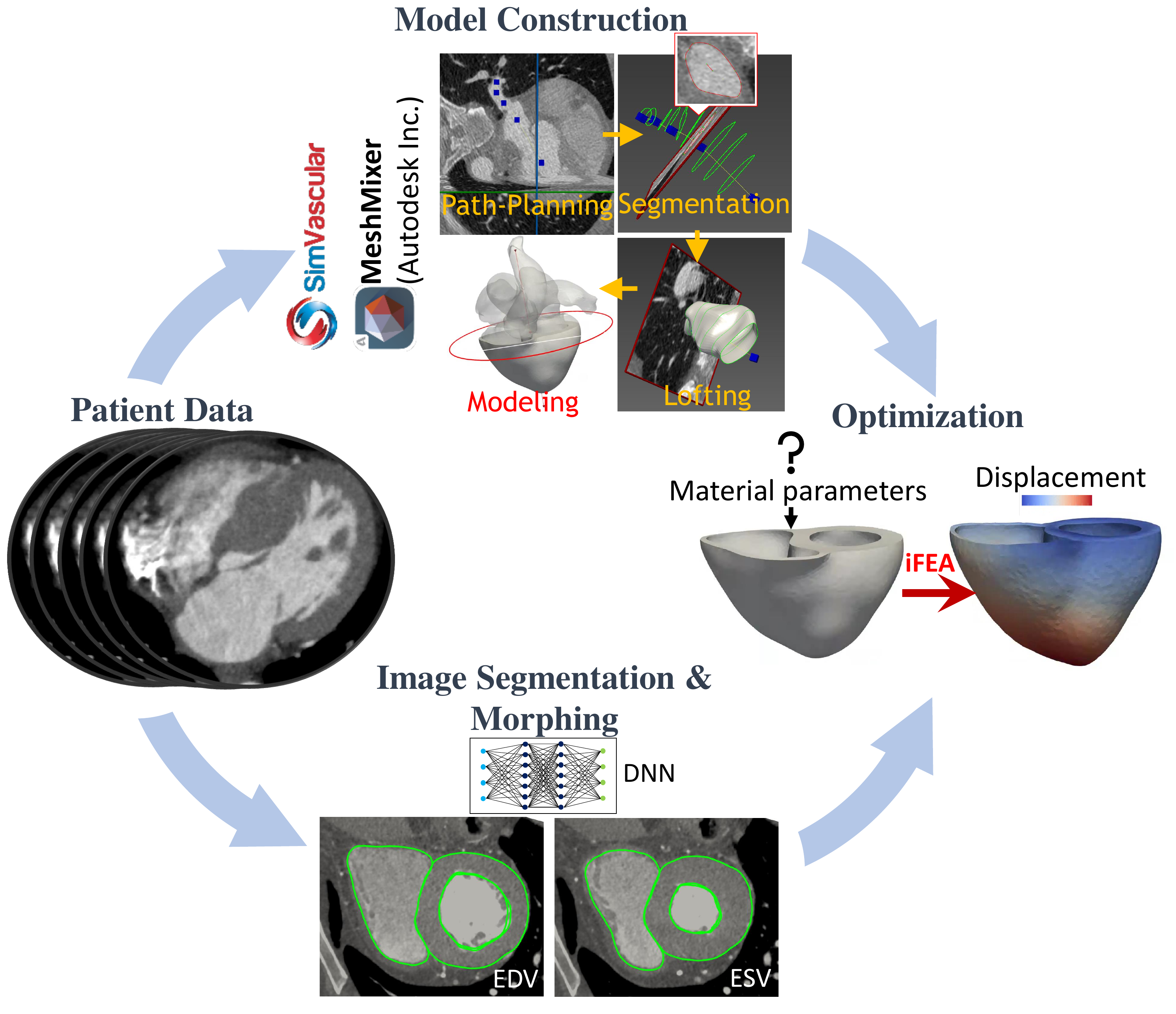}
\caption{Workflow for patient-specific modeling of myocardial mechanics. Time-dependent 3D-CT images are used as the starting point. The myocardium is segmented at different phases of the cardiac cycle, either manually using SimVascular and Meshixer (Autodesk Inc.),\cite{vedula2017method,baumler2020fluid} or using a machine learning approach based on deep neural networks (DNN).\cite{kong2023learning} An optimization framework is then applied to estimate the stress-free configuration and the passive material parameters specific to the patient's model. (iFEA: inverse finite element analysis; EDV: end-diastolic volume; ESV: end-systolic volume;)}
\label{fig_workflow}
\end{figure}

In our methodology, we construct models of the biventricular and LA myocardium from the patient’s time-resolved 3D CT data (Fig.~\ref{fig_workflow}). Subsequently, we employ a combination of automatic and manual segmentation methods to extract the myocardium and its motion during the cardiac cycle. Utilizing a rule-based myocardial fiber description,\cite{bayer2012novel,piersanti2021modeling} physiological boundary conditions,\cite{pfaller2019importance} and chamber pressures derived from literature and clinical evidence,\cite{klingensmith2008washington} we developed a nested optimization framework to simultaneously estimate the parameters governing passive myocardial mechanics and the reference configuration. We employ the orthotropic Holzapfel-Ogden (HO) and the transversely isotropic Guccione-McCulloch (GM) constitutive models to simulate the passive mechanical response of the myocardium.\cite{holzapfel2009constitutive,guccione1991passive} The optimization aims to minimize discrepancies in ventricular volumes and local nodal displacements between the FEA predictions and the corresponding data from medical images. This nested optimization scheme involves outer iterations optimizing the constitutive model parameters and inner iterations estimating the stress-free configuration. We employ different optimization methods in the outer iterations and compare their performance in estimating material parameters with robustness and accuracy. A sensitivity analysis is performed to evaluate the influence of mesh density, fiber orientation, and the location of these fiber directions in the solver on finding the best-fit material parameters and the cavity pressure-volume (P-V) profiles. We further assess the sensitivity of the P-V profiles to the material constitutive model parameters. Finally, we compare our approach against the Klotz EDPVR model, which relies on a single pressure-volume measurement and is typically used when time-resolved imaging data is unavailable.\cite{klotz2007computational}

The manuscript is structured as follows. Section~\ref{sec:methods} describes the patient characteristics and the proposed optimization framework (Fig.~\ref{fig_workflow}). In particular, we provide an overview of the image analysis, myocardial segmentation methods, and finite element (FE) model setup, highlighting the differences between biventricular and left atrial myocardial analysis. We review the notions of nonlinear elasticity, the constitutive models employed, and the stabilized variational multiscale (VMS) formulation employed to solve the governing elastodynamics equations. We finally describe the optimization framework that allows us to determine the stress-free configuration and the best-fit material parameters of a personalized passive myocardium. We provide the simulation results in Section~\ref{sec:results} for both normal and HOCM patients and discuss these results together with the sensitivity analysis in Section~\ref{sec:discussion}. Finally, we discuss model limitations and conclude in Sections~\ref{sec:limitation} and~\ref{sec:conclusion}, respectively.

%%%%%%%%%%%%%%%%%%%%%%%%%%%%%%%%%%%%%%%%%%%%%%%%%%%%%%%%%%%%%%%%%%%%%%%%%%%
%------------------------    METHODS   ------------------------------------
%%%%%%%%%%%%%%%%%%%%%%%%%%%%%%%%%%%%%%%%%%%%%%%%%%%%%%%%%%%%%%%%%%%%%%%%%%%
\section{Methods}\label{sec:methods}

\subsection{Patient data}\label{ssec:patient_data}

4D-CT images with a typical voxel resolution $0.43\cdot 0.43\cdot 0.5 \text{mm}^3$ ($512\cdot 512\cdot 280$ voxels) and about 20 volumes per cardiac cycle were obtained retrospectively for the control and three HOCM patients. While the control subject’s data was acquired from the Veterans Affairs Palo Alto Health Care System, the HOCM patients' data was obtained from the Columbia University Irving Medical Center, retrospectively. Data acquisition was conducted with prior approval from the respective Institutional Review Boards (IRB), and HIPAA-compliant procedures were followed throughout the study. A brief description of the clinical characteristics is provided in Table~\ref{table_demographics}. Further information on the data acquisition process and the imaging modality were reported for the chosen subjects elsewhere.\cite{chen2021preoperative,takayama2019virtual}

\begin{table}[ht]
\begin{center}
% \begin{minipage}{174pt}
\caption{Clinical characteristics of the subjects included in the study. HOCM: hypertrophic obstructive cardiomyopathy; BP: blood pressure in mmHg. HR: heart rate in beats per minute (bpm)}
\label{table_demographics}
\vspace{4pt}
\begin{tabular}{c c c c c}
\toprule
Subject & Age [y] &  Sex    & BP (high/low) & HR \\
\midrule
Control &   50    &  Male 	 &   98/53  & 87  \\
 HOCM1  &   34    &  Male   &   119/79 & 94  \\
 HOCM2  &   62    &  Male   &   150/82 & 65  \\
 HOCM3  &   41    &  Female &   107/64 & 69  \\
\bottomrule
\end{tabular}
% \end{minipage}
\end{center}
\end{table}

\subsection{Image processing \& segmentation}\label{ssec:img_seg_reg}

\begin{figure}[ht!]
\centering
\includegraphics[width=0.7\textwidth]{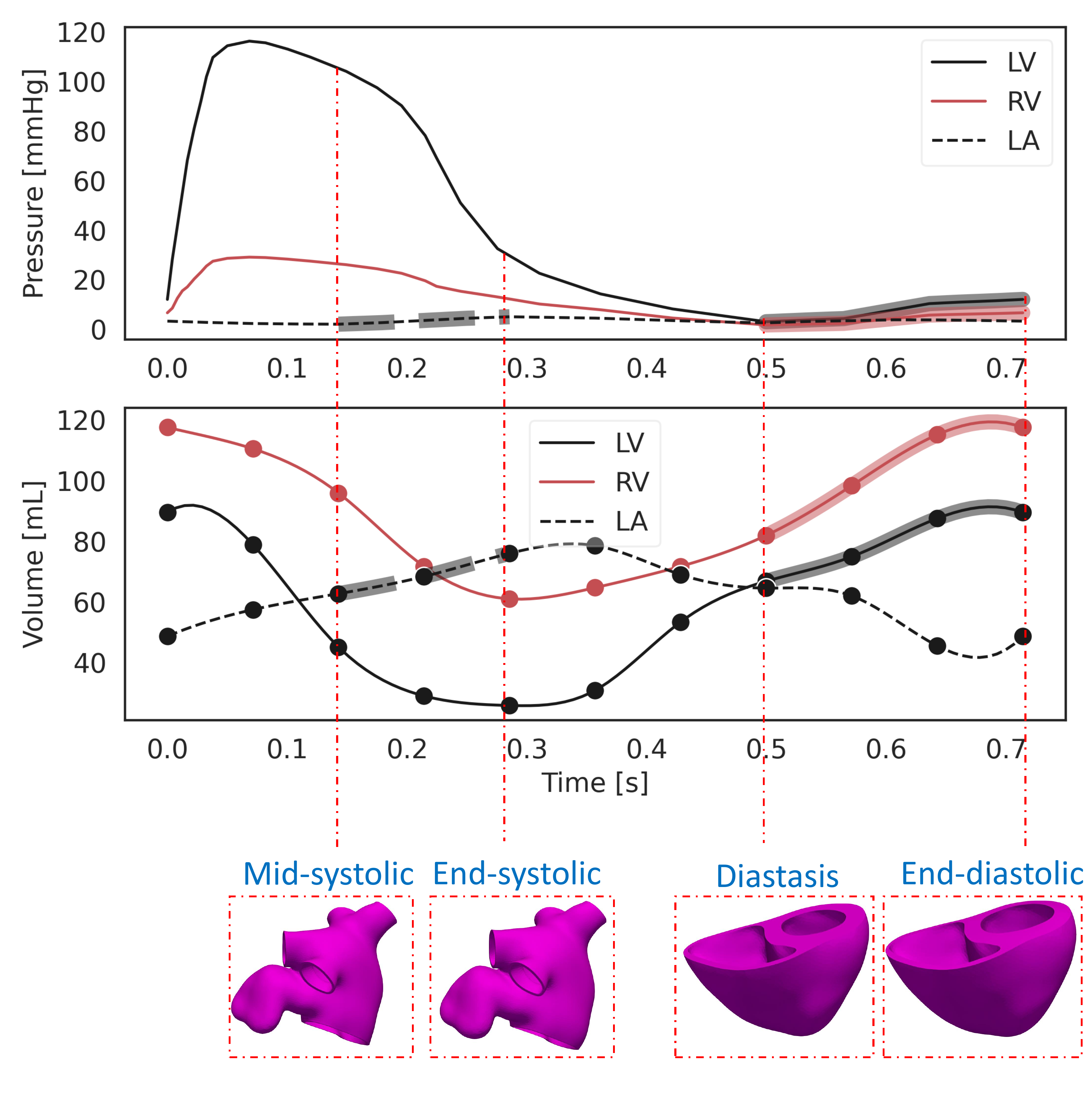}
\caption{Time-varying representative hemodynamic pressure (top) and cavity volumes (center) for the control subject. The pressure profiles are adapted from M.E. Klingensmith et al.\cite{klingensmith2008washington} The cavity volumes for each chamber are obtained by applying HeartDeformNet, a DNN-based automatic cardiac segmentation toolkit.\cite{kong2023learning} The manually segmented myocardium of the biventricle and left atrium (LA) is highlighted in the time interval of interest to simulate passive mechanics (bottom).}
\label{fig_image_registration}.
\end{figure}

In this study, we aim to simulate the passive mechanics of the ventricles and the left atrium (LA). We, therefore, focus on a select interval in the cardiac cycle when these chambers undergo a simple expansion under passively applied hemodynamic pressure. For the ventricles, this interval is during diastole, particularly the diastasis and late atrial filling (A-wave) phases. We exclude the early filling phase (E-wave), characterized by rapid ventricular relaxation following the preceding cycle's contraction (systole). However, for the left atrium, we focus on ventricular systole, during which the atrium is filled with blood from pulmonary veins while the mitral valve is closed. We also ignore the atrial relaxation due to elastic recoil from the preceding A-wave. Therefore, we segment the myocardium during late-diastole (70\%-100\% R-R\footnote[2]{Here, R-R denotes the interval between successive R-waves of the QRS complex on an electrocardiogram.}) for the biventricle and during late ventricular systole (20\%-40\% R-R) for LA (Fig. \ref{fig_image_registration}). These segmentations are used to extract changes in cavity volumes and local nodal displacements in the time interval of interest, which are then used in the objective function of the optimization algorithm. 

We used two different approaches for myocardial segmentation and motion extraction. We first applied HeartDeformNet, a deep neural network (DNN)-based whole-heart mesh reconstruction framework, to automatically segment the four chambers across all phases of the cardiac cycle.\cite{kong2023learning} While HeartDeformNet exhibited superior segmentation performance for the control subject, it failed to adequately capture the thick septum in HOCM patients due to the lack of HOCM data in its training dataset. Moreover, the right ventricular (RV) and LA myocardium were not segmented by HeartDeformNet as it is trained to extract only the lumen. Further, the left atrial appendage (LAA) was not segmented as LAA is not included in its training dataset as well. Therefore, we employed HeartDeformNet only for the control subject to extract (a) the cavity volumes of the two ventricles (Fig. \ref{fig_image_registration}) and (b) the local displacements of the LV myocardium and the endocardial surfaces of RV and LA.

For all other data needed for the optimization algorithm, we used our previously demonstrated manual segmentation workflow, using SimVascular\footnote{\url{https://simvascular.github.io/}} and Meshmixer\footnote{Autodesk Inc., \url{https://www.meshmixer.com/}, version: 3.5}.\cite{vedula2017method, baumler2020fluid} The manual approach involves creating paths and 2D segmentations and lofting the endocardial and epicardial surfaces in SimVascular.\cite{updegrove2017simvascular} Subsequent smoothing and decimation in Meshmixer refine the triangulated surfaces, following which a Boolean operation is performed to extract the thick myocardium. As the epicardial surface of the thin LA is not delineated on a CT scan, we simply extrude the LA endocardium by 2mm along the surface normal.\cite{piersanti2021modeling,hoermann2019automatic} However, when manual segmentation is performed at multiple phases, the surface mesh topology, including the mesh nodes and connectivity, is altered. Therefore, we employed a Bayesian Coherent Point Drift (BCPD) algorithm,\cite{hirose2020bayesian} implemented in the Python-based open-sourced tool, \emph{probreg\footnote{\url{https://github.com/neka-nat/probreg}}}, to deform the point set and establish a one-to-one correspondence between successive segmentations, thereby allowing us to compute nodal displacements. 

The manually segmented models include biventricular myocardium in the control subject at 70\% R-R, LA myocardium in the control subject for all cardiac phases, and the biventricular myocardium in HOCM patients for all cardiac phases. This dual approach for myocardial segmentation and motion extraction, combining machine learning and manual segmentation, was employed to obtain all the required data for the ensuing optimization step, including LA and biventricular cavity volumes and the myocardial local displacements for all the subjects considered in this study.

\begin{figure}[H]
\centering
\includegraphics[width=0.7\textwidth]{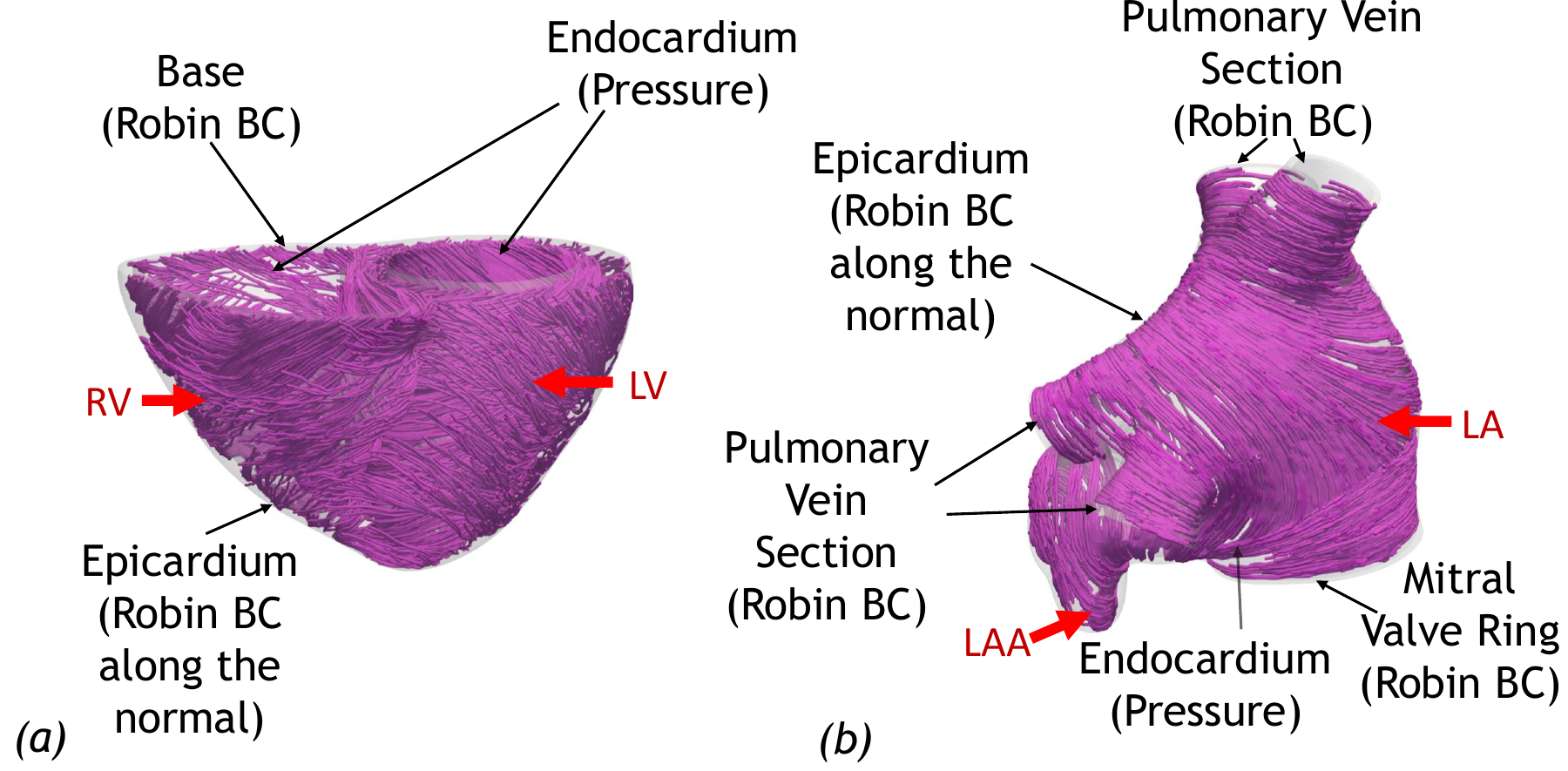}
\caption{Patient-specific myocardium models and fiber orientations of the (a) biventricle and (b) left atrium (LA) from a healthy subject, along with the boundary conditions used for the finite element simulations. For the biventricular model (a), hemodynamic pressure is applied on the left and right endocardial surfaces. Robin boundary conditions are applied on the basal plane and the epicardial surface to account for the effect of pericardium.\cite{pfaller2019importance} For the LA model (b), pressure is applied on the endocardial surface as a Neumann boundary condition, while Robin boundary conditions are applied on the pulmonary venous sections and the epicardial surface.}
\label{fig_fea_model}
\end{figure}

\subsection{Finite element model of the biventricle}\label{ssec:FE_BiV}

Patient-specific biventricular geometries (Fig.\ref{fig_fea_model}(a)) were derived from segmenting the volumetric CT data acquired during diastasis (70\% R-R). To simulate the effect of the pericardium, physiologically relevant Robin boundary conditions (Eq. (\ref{eq_bcs})) were applied to the surfaces of the FE model.\cite{pfaller2019importance} Given the lack of clinically measured pressures in our patient group, a generic pressure profile (Fig. \ref{fig_image_registration}, top) was applied to the LV and RV endocardial surfaces during diastole that matches the LV end-diastolic pressure of 12~mmHg and an RV end-diastolic pressure of 6.5~mmHg.\cite{klingensmith2008washington} The time duration from the diastasis to the end-diastolic state was set based on the subjects’ measured heart rate (Table~\ref{table_demographics}). Given that HOCM patients typically exhibit higher diastolic pressures, we scaled LV and RV pressures by 25\% of the control subject’s pressure values.\cite{nishimura2017hypertrophic} This simulation setup results in a physiologically observed ventricular deformation, where the apex undergoes minimal deformation while the basal plane moves predominantly along the apico-basal direction, contributing substantially to the chamber volume change.

The constitutive models employed in this study require specifying myofiber and sheet orientations within the myocardium (Fig.~\ref{fig_fea_model}). We adopted the Laplace-Dirichlet Rule-Based (LDRB) method to define the longitudinal and transverse fiber directions at the endocardial and epicardial surfaces.\cite{bayer2012novel} The longitudinal fibers ($\mathbf{f}_0$) vary from $40^{\circ}$ at the endocardium to $-50^{\circ}$ at the epicardium, where the fiber angle represents the helical angle measured with respect to the anti-clockwise circumferential direction when viewed from the basal plane towards the apex. Likewise, the transverse fiber direction or sheet angle ranged from $-65^{\circ}$ at the endocardium to $25^{\circ}$ at the epicardium, where the sheet angle was measured with respect to the local outward transmural axis, perpendicular to the longitudinal fiber direction.\cite{bayer2012novel} The sheet-normal vector ($\mathbf{n}_0$) is orthonormal to both the longitudinal and transverse fiber directions. We note that our definitions of the fiber $\mathbf{f}_0$, sheet $\mathbf{s}_0$, and sheet-normal $\mathbf{n}_0$ directions align with $F$, $T$, and $S$, respectively, as denoted in the original work by Bayer et al.\cite{bayer2012novel} We analyzed the sensitivity of fiber directions by varying the fiber and angles between $\pm20^{\circ}$ and $\pm60^{\circ}$ in increments of $20^{\circ}$ (Section~\ref{ssec:fiber_orient}). We also assessed the effect of storing the fiber and sheet directions at various levels within our finite element solver, including mesh nodes, element centroids, and Gauss quadrature points (Section~\ref{ssec:fiber_loc}).

\subsection{Finite element model of the left atrium}\label{ssec:FE_LA}

A patient-specific LA model is created from 3D-CT data of a healthy subject during mid-systole (20\% R-R) (Fig.\ref{fig_fea_model}(b)). Similar to the biventricular model, physiological boundary conditions are imposed on the LA surface for cardiac mechanics simulations.\cite{pfaller2019importance} Robin boundary conditions were applied on the distal sections of the pulmonary veins and the epicardium. The hemodynamic pressure in the LA cavity was assumed to increase from 2mmHg at 20\% R-R to 5~mmHg at 40\% R-R over a duration of 20\% of the cardiac cycle (Fig.~\ref{fig_image_registration}).\cite{klingensmith2008washington}

The mechanics of LA myocardium remains an active area of research.\cite{piersanti2021modeling,niederer2019computational,rossi2022rule} While various hyperelastic constitutive models have been employed in the past for the LA myocardium,\cite{pfaller2019importance,niederer2019computational,patelli2017isogeometric} a consensus has emerged among many researchers favoring a fiber-reinforced model.\cite{rossi2022rule,augustin2020impact} In this study, we opted for the orthotropic HO model, similar to the biventricular case. Following the procedure outlined in Piersanti et al.,\cite{piersanti2021modeling} we employed a rule-based method to define fiber directions on the LA myocardium. Specific threshold values for the left pulmonary vein, the right pulmonary vein, and the mitral valve are $\tau_\text{lpv}=0.65, \tau_\text{rpv}=0.1, \text{and } \tau_\text{mv}=0.65$, respectively.\cite{piersanti2021modeling}

%The FEA-ready models of the biventricle and LA myocardium are meshed with four-noded tetrahedral (TET4) elements using Tetgen\footnote{https://wias-berlin.de/software/tetgen/}, accessible within SimVascular\footnote{https://simvascular.github.io/}. A mesh convergence study was undertaken to obtain an optimal mesh size that yielded reasonable accuracy compared to the finest grid (Section \ref{ssec:mesh_size}).

\subsection{Passive mechanics formulation}\label{ssec:passive_mech}

We will establish the passive mechanics problem by concisely introducing the nonlinear continuum mechanics framework, the chosen constitutive models, and the stabilized variational multiscale (VMS) formulation for solving the governing equations of motion. For a deeper understanding of the theoretical and mathematical underpinnings, we refer interested readers to the comprehensive work by Liu et al.\cite{liu2018unified} Here, we use the biventricular case as a representative example, although extending it to modeling LA mechanics is straightforward.

\subsubsection{Kinematics, equations of motion, and boundary conditions}
Let $\pmb{\Omega_X}$ and $\pmb{\Omega_x}$ be bounded open sets in $\mathcal{R}^3$ with Lipschitz boundaries and represent the reference and current configurations, respectively. In this study, $\pmb{\Omega_X}$ denotes the stress-free reference configuration of the myocardium while $\pmb{\Omega_x}$ represents the deformed myocardium at any instant during passive expansion. We define the boundary, $\Gamma_\mathbf{x} := \Gamma_{\mathrm{base}}\cup\Gamma_{\mathrm{epi}}\cup\Gamma_{\mathrm{endo-lv}}\cup\Gamma_{\mathrm{endo-rv}}$, as the union of the basal plane ($\Gamma_{\mathrm{base}}$), epicardium ($\Gamma_{\mathrm{epi}}$), and endocardium of the left and the right ventricles ($\Gamma_{\mathrm{endo-lv}}$, $\Gamma_{\mathrm{endo-rv}}$). The motion of the body is then characterized by a deformation map, $\varphi : \pmb{\Omega_X} \rightarrow \pmb{\Omega_x}$ such that $\mathbf{x} = \varphi\left({\mathbf{X}}, t\right)$, where $\mathbf{x}$ is the current position of a material point at a time $t$ that was originally at $\mathbf{X}$ in the reference configuration. The displacement, $\mathbf{u}$ and velocity, $\mathbf{v}$ of a material particle are defined as,

\begin{equation}
\begin{split}
\mathbf{u} := & \varphi(\mathbf{X},t) - \mathbf{X} = \mathbf{x}(\mathbf{X},t) - \mathbf{X} \\
\mathbf{v} := & \frac{\partial \varphi}{\partial \mathbf{t}} \bigg|_\mathbf{X} = \frac{d \mathbf{u}}{dt}
\end{split}
\end{equation}

\noindent where, $d(\bullet)/dt$ is the total time derivative. The deformation gradient ($\mathbf{F}$), its Jacobian determinant ($J$), the right Cauchy-Green deformation tensor ($\mathbf{C}$), and the Green-Lagrange strain tensor ($\mathbf{E}$) are defined as,

\begin{equation}
\begin{split}
\mathbf{F} := \frac{\partial \varphi}{\partial\mathbf{X}}, \quad J := \det \big(\mathbf{F}\big), \quad \mathbf{C}=\mathbf{F}^T\mathbf{F}, \quad \mathbf{E}=\frac{1}{2}\big(\mathbf{C}- \mathbf{I}\big)
\end{split}
\end{equation}

\noindent We perform a multiplicative decomposition of the deformation gradient tensor to define $\overline{\mathbf{F}}$ and $\overline{\mathbf{C}}$ as,

\begin{equation}
\overline{\mathbf{F}} := J^{-1/3}\mathbf{F}, \quad \overline{\mathbf{C}} := J^{-2/3}\mathbf{C}
\end{equation}

\noindent that represent the splitting of local deformation ($\mathbf{F}$) into volume-preserving (isochoric, $\overline{\mathbf{F}}$) and volume-changing (dilatational, $J^{1/3}\mathbf{I}$) components. The mechanical behavior of the hyperelastic material is characterized by a Gibbs free energy $G(\overline{\mathbf{C}}, p)$, where $p$ is the thermodynamic pressure, and can be decoupled into isochoric ($G_\mathrm{iso}$) and volumetric components ($G_\mathrm{vol}$) as,\cite{liu2018unified}

\begin{equation}
G(\overline{\mathbf{C}},p) = G_\mathrm{iso} (\overline{\mathbf{C}}) + G_\mathrm{vol} (p)
\end{equation}

\noindent In the absence of body forces, the equations of the motion are then given as,

\begin{subequations}
\begin{align}
  \frac{d\mathbf{u}}{dt} - \mathbf{v} = & \ \mathbf{0} \quad \mathrm{in} \ \Omega_\mathbf{x} \label{eq_kin}\\
  \beta_{\theta}(p)\frac{dp}{dt} + \nabla_{\mathbf{x}}\cdot\mathbf{v} = & \ 0 \quad \mathrm{in} \ \Omega_\mathbf{x} \label{eq_mass} \\
  \rho(p)\frac{d\mathbf{v}}{dt} + \nabla_\mathbf{x} p - \nabla_\mathbf{x} \cdot \pmb{\sigma}_{dev} = & \ \mathbf{0} \quad \mathrm{in} \ \Omega_\mathbf{x} \label{eq_linMom}
\end{align}
\label{eq_govEqs}
\end{subequations}

\noindent where, $\beta_{\theta}$ is the isothermal compressibility coefficient, $\rho$ is the density of the material, $\pmb{\sigma}_{dev}$ represents the deviatoric part of the Cauchy stress tensor, and $\nabla_\mathbf{x}$ is the gradient operator defined in the spatial coordinates. In the above formulation, $p$ is an independent variable while $\rho$ and $\beta_{\theta}$ are dependent on $p$. Further, the first equation (Eq. (\ref{eq_kin})) represents the kinematic relation between the displacement and the velocity of the body, and the following two equations (Eqs. (\ref{eq_mass}, \ref{eq_linMom})) represent the conservation of mass and linear momentum, respectively. If $\rho_0$ is the density of the material in the reference configuration, the constitutive relations of the hyperelastic material are represented in terms of the specific Gibbs free energy components (per unit mass) as,\cite{liu2018unified}

\begin{subequations}
\begin{align}
    \rho(p) := \Bigg(\frac{\partial G_{vol}(p)}{\partial p}\Bigg)^{-1} \ , \quad \beta(p) := \frac{1}{\rho} \frac{d\rho}{dp} = & -\rho(p) \frac{\partial^2G_{vol}(p)}{\partial p^2 }, \label{eq_rho_beta} \\ 
    \pmb{\sigma}_{dev} := J^{-1}\overline{\mathbf{F}} \big(\mathbb{P}:\overline{\mathbf{S}}\big) \overline{\mathbf{F}}^{T} + 2\mu_v \mathrm{dev}[\mathbf{d}] \ , &\quad \overline{\mathbf{S}} = 2 \frac{\partial (\rho_0 G_{iso})}{\partial \overline{\mathbf{C}}} \label{eq_dev_stress} \\ 
    \pmb{\sigma} := \pmb{\sigma}_{dev} - p\mathbf{I} 
\end{align}
\label{eq_constRel}
\end{subequations}

\noindent where, $\mu_v$ is the dynamic shear viscosity, $\mathrm{dev}[\mathbf{d}]$ is the deviatoric part of the rate of deformation tensor, $\mathbf{d} := \frac{1}{2}\big(\nabla_\mathbf{x} \mathbf{v} + (\nabla_\mathbf{x} \mathbf{v})^T\big)$, and $\mathbb{P} = \mathbb{I} - \frac{1}{3}\big(\mathbf{C}^{-1} \otimes \mathbf{C}\big) $ is the projection tensor. The first term in Eq. (\ref{eq_dev_stress}) for $\pmb{\sigma}_{dev}$ represents the isochoric elastic stress, while the second term is the viscous shear stress.

We apply physiologically relevant boundary conditions on the myocardium as,\cite{pfaller2019importance}

\begin{equation}
\begin{split}
  \pmb{\sigma}\mathbf{\hat{n}} = & - p_{\mathrm{lv}}\mathbf{\hat{n}} \quad \mathrm{on} \ \Gamma_{\mathrm{endo-lv}} \\
  \pmb{\sigma}\mathbf{\hat{n}} = & - p_{\mathrm{rv}}\mathbf{\hat{n}} \quad \mathrm{on} \ \Gamma_{\mathrm{endo-rv}} \\
  \pmb{\sigma}\mathbf{\hat{n}} + \big(k_\mathrm{epi}\mathbf{u}\cdot\mathbf{\hat{n}} + c_\mathrm{epi}\mathbf{v}\cdot\mathbf{\hat{n}}\big)\mathbf{\hat{n}} = & -p_\mathrm{epi} \mathbf{\hat{n}} \quad \mathrm{on} \ \Gamma_{\mathrm{epi}} \\
  \pmb{\sigma}\mathbf{\hat{n}} + k_\mathrm{base}\mathbf{u} + c_\mathrm{base}\mathbf{v} = & \ \mathbf{0} \quad \mathrm{on} \ \Gamma_{\mathrm{base}}
\end{split}
\label{eq_bcs}
\end{equation}

\noindent where $k_{(\bullet)}$ and $c_{(\bullet)}$ are the stiffness and damping coefficients of the Robin boundaries, respectively, $p_\mathrm{lv}$ and $p_{\mathrm{rv}}$ are the hemodynamic pressures applied on the left and the right endocardium, and $p_\mathrm{epi}$ is the thoracic cavity pressure acting on the epicardium, which is typically small and is set to zero in this work. The equations (Eqs. (\ref{eq_govEqs}-\ref{eq_bcs})) complete the description of the initial-boundary value problem for the biventricular myocardial passive mechanics. 

For the left atrium, Robin boundary conditions are applied on the mitral valve ($\Gamma_{\mathrm{mv}}$) and pulmonary vein cross-sectional planes ($\Gamma_{\mathrm{pulv}}$), characterized by additional parameters $k_{\mathrm{mv}}, k_{\mathrm{pulv}}, c_{\mathrm{mv}}, \text{and } c_{\mathrm{pulv}}$. 

\begin{equation}
\begin{split}
  \pmb{\sigma}\mathbf{\hat{n}} = & - p_{\mathrm{lv}}\mathbf{\hat{n}} \quad \mathrm{on} \ \Gamma_{\mathrm{endo}} \\
  \pmb{\sigma}\mathbf{\hat{n}} + \big(k_\mathrm{epi}\mathbf{u}\cdot\mathbf{\hat{n}} + c_\mathrm{epi}\mathbf{v}\cdot\mathbf{\hat{n}}\big)\mathbf{\hat{n}} = & -p_\mathrm{epi} \mathbf{\hat{n}} \quad \mathrm{on} \ \Gamma_{\mathrm{epi}} \\
  \pmb{\sigma}\mathbf{\hat{n}} + k_\mathrm{mv}\mathbf{u} + c_\mathrm{mv}\mathbf{v} = & \ \mathbf{0} \quad \mathrm{on} \ \Gamma_{\mathrm{mv}} \\
  \pmb{\sigma}\mathbf{\hat{n}} + k_\mathrm{pulv}\mathbf{u} + c_\mathrm{pulv}\mathbf{v} = & \ \mathbf{0} \quad \mathrm{on} \ \Gamma_{\mathrm{pulv}}
\end{split}
\label{eq_la_bcs}
\end{equation}

\subsubsection{Constitutive models}\label{ssec:const_models}

We model the myocardium as a hyperelastic, nearly-incompressible material, using two widely-used anisotropic constitutive models, the orthotropic modified Holzapfel-Ogden (HO) model,\cite{holzapfel2009constitutive, nolan2014robust} and the transversely-isotropic Guccione-McCulloch (GM) model.\cite{guccione1991passive} The isochoric components of the corresponding Gibbs free energies are given as,

\begin{equation}
\begin{split}
    G_\mathrm{iso}^{\mathrm{HO}}(\overline{\mathbf{C}}) & = \frac{a^\mathrm{HO}}{2b^\mathrm{HO}}\exp \{b^\mathrm{HO}(\overline{I}_1 - 3)\} \\
    & + \sum_{i\in {f,s}} \frac{a_i^\mathrm{HO}}{2b_i^\mathrm{HO}}\chi(I_{4,i})\bigg(\exp \{b_i^\mathrm{HO} (I_{4,i} - 1 )^2\} - 1\bigg) \\
    &  +  \frac{a_{fs}^\mathrm{HO}}{2b_{fs}^\mathrm{HO}} \bigg(\exp \{b_{fs}^\mathrm{HO}I^2_{8,fs}\} - 1\bigg) 
\end{split}
\label{eq_const_ho}
\end{equation}

\begin{equation}
\begin{split}
    G_\mathrm{iso}^{\mathrm{GM}}(\overline{\mathbf{E}}^{\ast}) & = \frac{C^\mathrm{GM}}{2}\bigg(\exp\{Q(\overline{\mathbf{E}}^{\ast})\} - 1\bigg) \ , \mathrm{where} \\
    Q(\overline{\mathbf{E}}^{\ast}) & = b_f^\mathrm{GM} {\overline{E}^\ast_{ff}}^2 + b_s^\mathrm{GM}\big({\overline{E}^\ast_{ss}}^2 + {\overline{E}^\ast_{nn}}^2 + 2{\overline{E}^\ast_{sn}}^2\big) \\
    & + 2b_{fs}^\mathrm{GM}\big({\overline{E}^\ast_{fs}}^2 + {\overline{E}^\ast_{fn}}^2\big)
\end{split}
\label{eq_const_gm}
\end{equation}

\noindent where $\overline{\mathbf{E}}^{\ast}$ is the local Green-Lagrange strain tensor.\footnote[2]{If $\mathbf{R}$ is the rotation matrix that transforms from local orthogonal coordinates parallel to $[\mathbf{f_0} \ \mathbf{s_0} \  \mathbf{n_0}]$ to the global Cartesian coordinates, then $\overline{\mathbf{E}}^{\ast}$ is related to its global counterpart, $\overline{\mathbf{E}} := \frac{1}{2}\big(\overline{\mathbf{C}}-\mathbf{I}\big)$, as $\overline{\mathbf{E}}^{\ast} = \mathbf{R}^T \overline{\mathbf{E}} \mathbf{R}$.} We also incorporate a smooth approximation of the Heaviside function, $\displaystyle \chi(\eta) := \frac{1}{1+\exp [-k_\chi (\eta - 1)]}$, to avoid any numerical instabilities under small compressive strains.\cite{pfaller2019importance}

In Eq. (\ref{eq_const_ho}), the set $\{a^\mathrm{HO},\ b^\mathrm{HO},\ a_f^\mathrm{HO},\ b_f^\mathrm{HO},\ a_s^\mathrm{HO},\ b_s^\mathrm{HO},\ a_{fs}^\mathrm{HO},\ b_{fs}^\mathrm{HO}\}$ defines the parameters for the HO model, while the set $\{C^\mathrm{GM},\ b_f^\mathrm{GM},\ b_s^\mathrm{GM},\ b_{fs}^\mathrm{GM}\}$ in Eq. (\ref{eq_const_gm}) defines the parameters for the GM model. Further, $\overline{I}_1$ ($:=\mathrm{tr}(\overline{\mathbf{C}})$) is the isotropic invariant that captures isochoric deformations, $I_{4f}$ ($:= \mathbf{f}_0\cdot \mathbf{C}\mathbf{f}_0$) and $I_{4s}$ ($:= \mathbf{s}_0\cdot \mathbf{C}\mathbf{s}_0$) are the transverse invariants for the fiber and sheet directions, respectively, and $I_{8fs}$ ($:= \mathbf{f}_0\cdot \mathbf{C}\mathbf{s}_0$) is the anisotropic invariant that captures the fiber-sheet interactions. 

\begin{remark}
Although $I_{8fs}$ is included in the definition of the free energy ($G_\mathrm{iso}^{\mathrm{HO}}$), it doesn’t contribute to the stress and stiffness tensors for the myocardial mechanics, because the fiber and sheet directions computed using the LDRB method are orthogonal to each other (Section \ref{ssec:FE_BiV}).
\end{remark}

\begin{remark}
While defining $G_\mathrm{iso}^{\mathrm{HO}}$, we use isochoric invariant ($\overline{I}_1$) for the isotropic term but use full invariants ($I_{4f}, I_{4s}, I_{8fs}$) for the anisotropic terms to correctly capture the volumetric anisotropic deformations.\cite{nolan2014robust}
\end{remark}

To model the near-incompressibility of the myocardium, we employ the widely used ST91 volumetric strain energy model,\cite{simo1991quasi} given in the form of specific Gibbs free energy as,

\begin{equation}
    G_{\mathrm{vol}}(p)\bigg\vert^{ST91} = \frac{-p^2 + p\sqrt{p^2 + \kappa^2}}{2 \kappa \rho_0} - \frac{\kappa}{2\rho_0} \ln \bigg( \frac{\sqrt{p^2 + \kappa^2} - p}{\kappa} \bigg)
\label{eq_ST91_Gvol}
\end{equation}

\noindent where $\kappa$ is the bulk modulus and $\rho_0$ is the density in the reference configuration. The corresponding $\rho(p)$ and $\beta_{\theta}(p)$ (Eq. \ref{eq_rho_beta}) are given by,

\begin{equation}
    \rho (p)\bigg\vert^\mathrm{ST91} = \frac{\rho_0}{\kappa} \Big( \sqrt{p^2 + \kappa^2} + p \Big) \ , \ \ \beta_{\theta}(p) \bigg\vert^\mathrm{ST91} = \frac{1}{\sqrt{p^2 + \kappa^2}}
\label{eq_ST91_rho}
\end{equation}

\subsubsection{Finite Element Discretization}\label{ssec:fea_vms}

The FEA-ready models of the biventricle (Section~\ref{ssec:FE_BiV}) and LA myocardium (Section~\ref{ssec:FE_LA}) are meshed with four-noded tetrahedral (TET4) elements using Tetgen\footnote{https://wias-berlin.de/software/tetgen/}, accessible within SimVascular\footnote{https://simvascular.github.io/}. A mesh convergence study was undertaken to obtain an optimal mesh size that yielded reasonable accuracy compared to the finest grid (Section \ref{ssec:mesh_size}). Our choice of TET4 elements was motivated by their versatility and robustness in meshing complex shapes, such as patient-specific geometries derived from medical images. However, resolving finite, incompressible deformations with linear tetrahedrons is not straightforward and often leads to volumetric locking issues.\cite{hughes2012finite}

In our current mixed formulation involving velocity ($\mathbf{v}$) and pressure ($p$) as the unknown degrees of freedom (see Remark below), only a specific choice of discrete function spaces satisfying the \emph{inf-sup} conditions yield meaningful solutions.\cite{hughes2012finite} While several accurate and robust methods have been developed to overcome volumetric locking issues in incompressible elasticity and plasticity, such as the B-bar and F-bar methods,\cite{elguedj2008b} selective and reduced integration methods,\cite{malkus1978mixed} their applicability is often limited to hexahedral meshes or other high-order elements. Generating such meshes for arbitrary patient-specific myocardium poses significant challenges.

\begin{remark}
In principle, we have nodal displacements ($\mathbf{u}$), velocity ($\mathbf{v}$), and pressure ($p$) as the unknown degrees of freedom (Eq.~(\ref{eq_govEqs})). However, $\mathbf{u}$ can be decoupled from the system of equations via a segregated two-step algorithm that effectively reduces the problem size by solving first for $\big\{\mathbf{v},p\big\}$, followed by a simple nodal update to obtain $\mathbf{u}$.\cite{liu2018unified,rossi2016implicit}
\end{remark}

To address these issues, we adopt stabilized finite elements based on the variational multiscale (VMS) formulation to spatially discretize and solve the governing equations (Eqs. (\ref{eq_govEqs}-\ref{eq_bcs})). The VMS formulation allows us to employ equal-order interpolation for velocity and pressure basis functions (P1-P1 for TET4), circumventing the \emph{inf-sup} conditions while mitigating volumetric locking, particularly in near-incompressible limit.\cite{liu2018unified, rossi2016implicit, karabelas2022accurate} 

We will now briefly discuss the semi-discrete form of the balance laws (Eq.~(\ref{eq_govEqs})) using VMS formulation. Let $\pmb{\Omega_x}$ be partitioned into $n_{el}$ non-overlapping elements $\pmb{\Omega_x}^e$, with $h^e$ being an element's characteristic size. The finite element trial solution spaces, $\mathcal{S}_{\mathbf{u}_h}$, $\mathcal{S}_{\mathbf{v}_h}$, and $\mathcal{S}_{p_h}$ for $\mathbf{u}$, $\mathbf{v}$, and $p$ are defined as,

\begin{equation}
\begin{split}
    \mathcal{S}_{\mathbf{u}_h} = & \Big\{ \mathbf{u}_h | \mathbf{u}_h (\mathbf{x}, t) \in (H^1)^{n_{sd}} \times [0, T], \mathbf{u}_h(\cdot, t) = \mathbf{g} \mathrm{~on~} \Gamma^g_{\mathbf{x}} \Big\} \\ 
    \mathcal{S}_{\mathbf{v}_h} = & \Big\{ \mathbf{v}_h | \mathbf{v}_h (\mathbf{x}, t) \in (H^1)^{n_{sd}} \times [0, T], \mathbf{v}_h(\cdot, t) = \frac{d \mathbf{g}}{d t} \mathrm{~on~} \Gamma^g_{\mathbf{x}} \Big\} \\ 
    \mathcal{S}_{p_h} = & \Big\{ p_h | p_h (\mathbf{x}, t) \in L^2 \times [0, T] \Big\}
\end{split}
\label{eq_fn_trial}
\end{equation}  

\noindent and the corresponding test function spaces are defined as,

\begin{equation}
\begin{split}
    \mathcal{V}_{\mathbf{u}_h} = & \Big\{ \mathbf{w}_{\mathbf{u}_h} | \mathbf{w}_{\mathbf{u}_h} (\mathbf{x}) \in (H^1)^{n_{sd}}, \mathbf{w}_{\mathbf{u}_h} = \mathbf{0} \mathrm{~on~} \Gamma^g_{\mathbf{x}} \Big\} \\ 
    \mathcal{V}_{\mathbf{v}_h} = & \Big\{ \mathbf{w}_{\mathbf{v}_h} | \mathbf{w}_{\mathbf{v}_h} (\mathbf{x}) \in (H^1)^{n_{sd}}, \mathbf{w}_{\mathbf{v}_h} = \mathbf{0} \mathrm{~on~} \Gamma^g_{\mathbf{x}} \Big\} \\ 
    \mathcal{V}_{p_h} = & \Big\{ w_{p_h} | w_{p_h} (\mathbf{x}) \in L^2 \Big\} \\ 
\end{split}
\label{eq_fn_test}
\end{equation}  

\noindent where $n_{sd}$ is the number of spatial dimensions, $T$ is a suitable time interval of interest, and $\mathbf{g}$ is the prescribed displacement on the Dirichlet boundary $\Gamma^g_{\mathbf{x}}$. The semi-discrete problem is then stated as,

Find $\mathbf{u}(t) \in \mathcal{S}_{\mathbf{u}_h}$, $\mathbf{v}(t) \in \mathcal{S}_{\mathbf{v}_h}$, and $p(t) \in \mathcal{S}_{p_h}$, such that for $t \in [0,T]$ and $\mathbf{y}:=\big\{\mathbf{u}, \mathbf{v}, p\big\}^T$,

\begin{equation}
\begin{split}
    \mathbf{0} = \mathbf{B}_k (\mathbf{w}_{\mathbf{u}} ; \dot{\mathbf{y}}, \mathbf{y}) & := \int_{\pmb{\Omega_x}} \mathbf{w}_{\mathbf{u}} \cdot \bigg( \frac{d \mathbf{u}}{d t} - \mathbf{v}\bigg) \ d \Omega_\mathbf{x}   \\
    \mathbf{0} = \mathbf{B}_m (\mathbf{w}_{\mathbf{v}} ; \dot{\mathbf{y}}, \mathbf{y}) & := \int_{\pmb{\Omega_x}} \bigg( \mathbf{w}_{\mathbf{v}} \cdot \rho (p) \frac{d \mathbf{v}}{d t} - \nabla_{\mathbf{x}} \cdot \mathbf{w}_{\mathbf{v}} p + \nabla_{\mathbf{x}} \mathbf{w}_{\mathbf{v}} : \pmb{\sigma}_{dev} \bigg) \ d \Omega_\mathbf{x} \\
    & - \int_{\Gamma_{\mathbf{x}}} \mathbf{w}_{\mathbf{v}} \cdot \mathbf{h} \ d \Gamma_{\mathbf{x}} + \sum_{e=1}^{n_{el}} \int_{\pmb{\Omega_x}^e} \nabla_{\mathbf{x}} \cdot \mathbf{w}_{\mathbf{v}} \pmb{\tau}_c r_c \ d \Omega_{\mathbf{x}^e} \\ 
    0 = \mathbf{B}_c (w_{p} ; \dot{\mathbf{y}}, \mathbf{y}) & := \int_{\pmb{\Omega_x}} w_{p} r_c \ d \Omega_{\mathbf{x}} + \sum_{e=1}^{n_{el}} \int_{\pmb{\Omega_x}^e} \nabla_{\mathbf{x}} w_{p_h} \cdot \pmb{\tau}_m \mathbf{r}_m \  d \Omega_{\mathbf{x}^e}
\end{split}
\label{eq_weak_VMS}
\end{equation}

\noindent where, $\mathbf{h}$ is the surface traction (Eqs.~(\ref{eq_bcs}-\ref{eq_la_bcs})), $\mathbf{B}_k$, $\mathbf{B}_m$, and $\mathbf{B}_c$ represent the usual bilinear forms of the kinematic constraint and the linear momentum and mass conservation equations (Eq.(~\ref{eq_govEqs})), respectively. The terms under summation represent the VMS-related stabilization terms, proportional to the residues in the continuity ($r_c$) and momentum ($\mathbf{r}_m$) equations, and controlled by the stabilization coefficients, $\pmb{\tau}_c$ and $\pmb{\tau}_m$, respectively. These are defined as,

\begin{equation}
\begin{split}
    r_c & = \beta_\theta (p) \frac{d p}{d t} + \nabla_\mathbf{x} \cdot \mathbf{v} \\
    \tau_c & = c_{\tau_c} h^e c \rho \\
    \mathbf{r}_m & = \rho(p) \frac{d \mathbf{v}}{d t} + \nabla_{\mathbf{x}} p - \nabla_{\mathbf{x}} \cdot \pmb{\sigma}_{dev} \\
    \pmb{\tau}_m & = \tau_m \mathbf{I}_{n_{sd}}\ , \tau_m = c_{\tau_m} \frac{h^e}{c\rho}
\end{split}
\label{eq_res_stab}
\end{equation}

\noindent where, $c_{\tau_c}$ and $c_{\tau_m}$ are the stabilization parameters, $h^e$ is the characteristic element size, proportional to the element Jacobian $\big(h^e = 0.5(J_e)^{1/n_{sd}}\big)$, and $c$ is the maximum wave speed in the structure,

\begin{equation}
\begin{split}
    c = & \sqrt{\frac{\lambda+ 2\mu}{\rho_0}} \text{ for incompressible materials, and} \\ 
    c = & \sqrt{\frac{\mu}{\rho_0}} \text{ for compressible materials} 
\end{split}
\label{eq_wave_speed}
\end{equation} 

\noindent where $\lambda$ and $\mu$ are the Lam\'e parameters of the isotropic linear elastic material, and $\rho_0$ is the density in the reference configuration. In the current work, we set $c_{\tau_c} = c_{\tau_m} = 10^{-5}$.

\subsection{Numerical Procedure}\label{ssec:num_proc}

Simulations were performed using an in-house multiphysics finite element solver, adapted from the open-source finite element solver, \emph{svFSI}\footnote{\url{https://github.com/SimVascular/svFSI}},\cite{zhu2022svfsi} which was previously validated and employed for various cardiovascular biomechanics applications, including simulating cardiac electrophysiology,\cite{salvador2024branched} myocardial mechanics,\cite{brown2024modular} blood flow in coronaries,\cite{seo2020effects, khan2021low} developing ventricles,\cite{vedula2017method, lee2013moving} and fluid-structure interaction (FSI) modeling in aortic dissection and aneurysms.\cite{baumler2020fluid, wang2022fluid, bazzi2022experimental} We also verified the solver’s capabilities for cardiac mechanics applications using the benchmark problems proposed in Land et al.\cite{land2015verification} Relevant data is shared in \ref{sec:suppl_verify}. We solved the nonlinear system of equations using the Newton-Raphson method, embedded within a segregated, two-stage predictor-multi-corrector algorithm, and integrated in time using the generalized-$\alpha$ method.\cite{liu2018unified} Within each nonlinear iteration, a block matrix equation of the type $\mathbf{A} \mathbf{x} = \mathbf{B}$ is solved. In all the simulations performed in this study, we employed the iterative solver, generalized minimal residual (GMRES), to solve the sparse system of linear equations.\cite{saad1986gmres} The linear solver convergence was accelerated using the additive Schwarz method (ASM) preconditioner included in the PETSc package\footnote{\url{https://github.com/petsc}}.\cite{dryja1991additive} For all the simulations, we set the tolerances for the nonlinear and the linear solvers to $10^{-6}$. We further set the spectral radius of infinite time step ($\rho_\infty$) in the generalized-$\alpha$ method to 0.5, which guarantees $2^{nd}$-order accuracy while ensuring optimal dissipation of high-frequency temporal oscillations.

\begin{figure}[ht]
\centering
\includegraphics[width=0.9\textwidth]{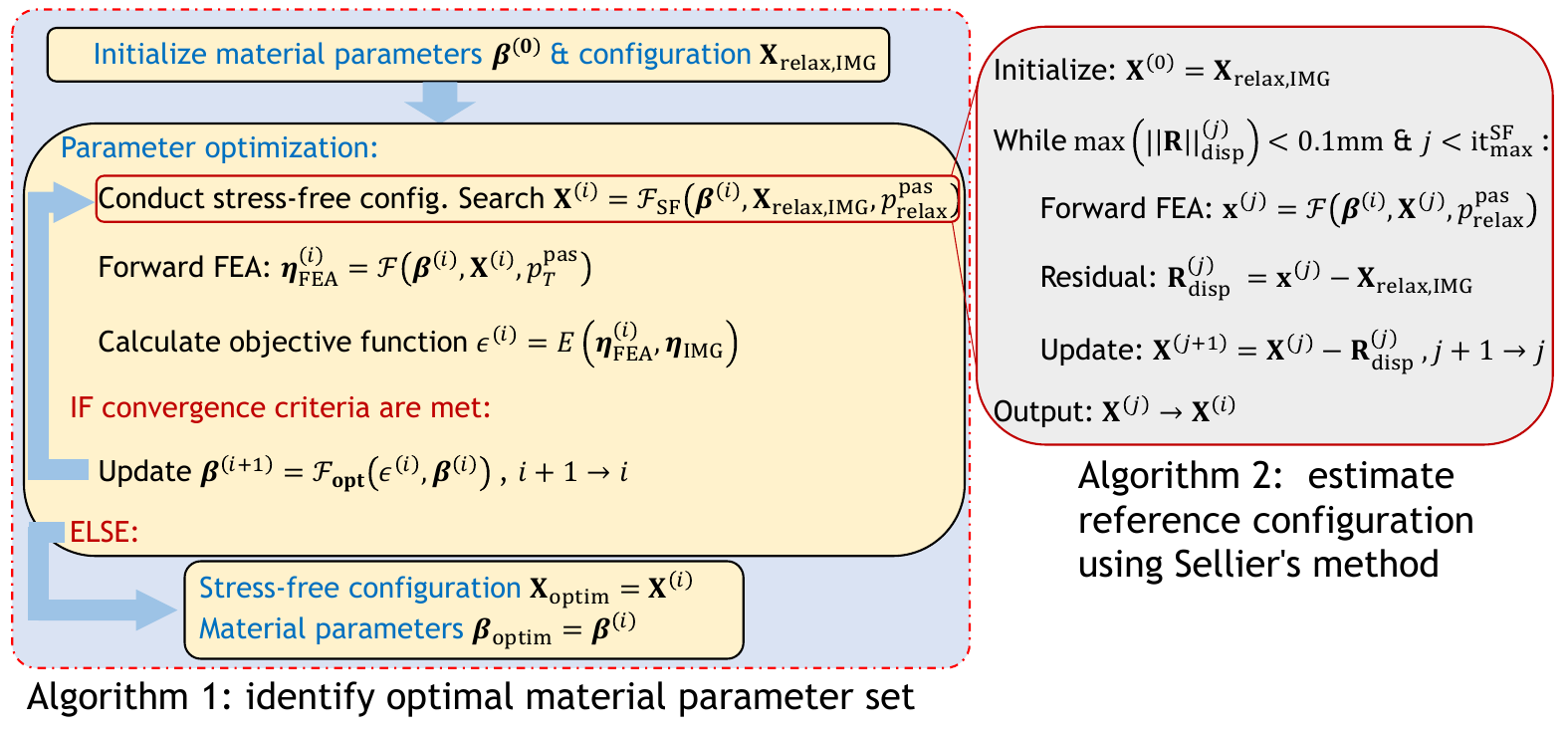}
\caption{Inverse FEA using a nested optimization scheme to estimate the stress-free configuration (inner loop) using augmented Sellier's method and the best-fit material parameters (outer loop).}
\label{fig_ifea_algo}
\end{figure}

\subsection{Inverse FEA using nested optimization}\label{ssec:ifea}

Our objective is to determine the material parameters that most accurately match the FEA-predicted myocardial deformation with the image-based motion extracted at multiple cardiac phases. However, these FE models are constructed from image data acquired under \textit{in vivo} conditions when the myocardium is stressed under hemodynamic pressure. Therefore, we must also identify the stress-free configuration that represents the topology of the myocardium under net zero hemodynamic load, a vital component of the nonlinear continuum mechanics framework.

This problem can be addressed in multiple ways. One could opt for a comprehensive optimization strategy that involves adjusting the material parameters, while simultaneously seeking the reference configuration, for each pressure load corresponding to the cardiac phase. However, we found that, in practice, this approach often leads to convergence issues and incurs substantial computational costs.

We, therefore, introduce a nested multi-level optimization method in which estimating the stress-free configuration is nested within an outer iterative loop, searching for optimal material parameters (Fig.~\ref{fig_ifea_algo}). At the outer level (Algorithm 1 in Fig.~\ref{fig_ifea_algo}), we iterate on the material parameters using a traditional optimization method to best approximate the deformed configuration with the image data at varying pressures during passive expansion. Meanwhile, the inner iterations (Algorithm 2 in Fig.~\ref{fig_ifea_algo}) employ augmented Sellier's algorithm to determine the stress-free reference configuration, for a given set of material parameters, set by the outer level.\cite{marx2022robust,sellier2011iterative}

\begin{table}[ht]
\begin{center}
% \begin{minipage}{174pt}
\footnotesize
\caption{Fixed parameters in the inverse FEA of biventricular myocardium}
\label{table_fix_params_biv}
\vspace{4pt}
\begin{tabular}{c c c c c c c c c}
\toprule
$\rho_0$  & $\kappa$ & $\mu_v$ & $k_{\mathrm{base}}$ & $c_{\mathrm{base}}$ & $k_{\mathrm{epi}}$ & $c_{\mathrm{epi}}$    &    $a_{fs}^{\mathrm{HO}}$  & $b_{fs}^{\mathrm{HO}}$   \\
%\midrule
\arrayrulecolor{black!30}\midrule
$1.055$  & $10^6$  & $10^3$ & $10^4$ & $500$ & $10^4$  & $500$ & $2160$ & $11.436$  \\
\arrayrulecolor{black}\bottomrule
\end{tabular}
\end{center}
\begin{tablenotes}
    \centering
    \vspace{-3pt}
    \footnotesize
    %\scriptsize
    \item $\rho_0$: $\text{g}/\text{cm}^3$; $\kappa$: $\text{dyn}/\text{cm}^2$; $\mu_v$: $\text{dyn-s}/\text{cm}^{2}$; $k_{(\cdot)}$: $\text{dyn}/\text{cm}^3$; 
    \item $c_{(\cdot)}$: $\text{dyn-s}/\text{cm}^3$; $a_{fs}^{\mathrm{HO}}$: $\text{dyn}/\text{cm}^2$; $b_{fs}^{\mathrm{HO}}$ is dimensionless.
\end{tablenotes}
\end{table}

\vspace{-3pt}
\begin{table}[H]
\begin{center}
% \begin{minipage}{174pt}
\footnotesize
\caption{Fixed parameters in the inverse FEA of LA myocardium}
\label{table_fix_params_la}
\vspace{4pt}
\begin{tabular}{c c c c c c c c c c c}
\toprule
$\rho_0$ & $\kappa$ & $\mu_v$ & $k_{\mathrm{mv}}$ & $c_{\mathrm{mv}}$ & $k_{\mathrm{epi}}$ & $c_{\mathrm{epi}}$   & $k_{\mathrm{pulv}}$ & $c_{\mathrm{pulv}}$ & $a_{fs}^{\mathrm{HO}}$ & $b_{fs}^{\mathrm{HO}}$   \\
\arrayrulecolor{black!30}\midrule
$1.055$ & $10^6$ & $10^3$ & $10^4 $ & $500$ & $10^4$ & $500$ & $10^6 $ & $500$ & $2160$ & 11.436  \\
\arrayrulecolor{black}\bottomrule
\end{tabular}
\end{center}
\begin{tablenotes}
    \centering
    \vspace{-3pt}
    \footnotesize
    %\scriptsize
    \item $\rho_0$: $\text{g}/\text{cm}^3$; $\kappa$: $\text{dyn}/\text{cm}^2$; $\mu_v$: $\text{dyn-s}/\text{cm}^{2}$; $k_{(\cdot)}$: $\text{dyn}/\text{cm}^3$; 
    \item $c_{(\cdot)}$: $\text{dyn-s}/\text{cm}^3$; $a_{fs}^{\mathrm{HO}}$: $\text{dyn}/\text{cm}^2$; $b_{fs}^{\mathrm{HO}}$ is dimensionless.
\end{tablenotes}
\end{table}

\vspace{-3pt}
\begin{table}[H]
\begin{center}
% \begin{minipage}{174pt}
\footnotesize
\caption{Constitutive model parameter bounds for optimization.}
\label{table_param_bounds}
\vspace{4pt}
\begin{tabular}{c c c c c c}
\toprule
$a^{\mathrm{HO}}$ & $b^{\mathrm{HO}}$ & $a_f^{\mathrm{HO}}$  & $b_f^{\mathrm{HO}}$ & $a_s^{\mathrm{HO}}$ & $b_s^{\mathrm{HO}}$ \\
\arrayrulecolor{black!30}\midrule
$ 10^2-10^4$ & $1-50$ & $10^3-10^6 $ & $1-50$ & $10^2-10^5 $ & $1-50$  \\
\arrayrulecolor{black}\bottomrule
\end{tabular}
\begin{tabular}{c c c c}
$C^{\mathrm{GM}}$ & $b_f^{\mathrm{GM}}$ & $b_s^{\mathrm{GM}}$ & $b_{fs}^{\mathrm{GM}}$ \\
\arrayrulecolor{black!30}\midrule
$10^1-10^4$ & $1-500$  & $1-100$ & $1-100$ \\
\arrayrulecolor{black}\bottomrule
\end{tabular}
% \end{minipage}
\end{center}
\begin{tablenotes}
    \centering
    \vspace{-3pt}
    \footnotesize
    %\scriptsize
    \item $a^{\mathrm{HO}}_{(\cdot)}$: $\text{dyn}/\text{cm}^2$; $C^{\mathrm{GM}}$: $\text{dyn}/\text{cm}^2$. $b_{(\cdot)}$ is dimensionless.
\end{tablenotes}
\end{table}

Throughout the optimization, our algorithm iterates on the parameters associated with the isochoric constitutive models (HO, GM), as identified in Eqs. (\ref{eq_const_ho},~\ref{eq_const_gm}), within a predefined set of bounds (Table~\ref{table_param_bounds}). All other parameters remain fixed (Tables~\ref{table_fix_params_biv},~\ref{table_fix_params_la}), and most of these are extracted from Pfaller et al.\cite{pfaller2019importance} The stiffness and damping coefficients on the basal plane ($k_{\text{base}}, c_{\text{base}}$) of the biventricular myocardium are set to be the same as the epicardium. The parameter set ($a_{fs}, b_{fs}$), governing the fiber-sheet interactions due to the anisotropic invariant term ($I_{8fs}$) in the HO model (Eq. (\ref{eq_const_ho})), is unchanged during optimization (see Remark in Section \ref{ssec:const_models}). Lastly, the HO model's Heaviside function parameter ($k_{\chi}$) was assigned a constant value of 100 and remains unaffected during optimization.\cite{pfaller2019importance}

We will now elaborate on the optimization algorithm in detail. It is convenient to define a starting point, i.e., the diastasis phase ($\sim$70\% R-R) for the biventricle or the mid-systolic phase ($\sim$20\% R-R) for the left atrium, as the relaxed state. The optimization begins with our initially segmented FEA model at the relaxed state, $\mathbf{X}_{\text{relax,IMG}}$, the corresponding fiber directions created using the LDRB method, and an initial set of material parameters ($\pmb{\beta}^{(0)}$). 

\begin{figure}[ht]
\centering
\includegraphics[width=0.85\textwidth]{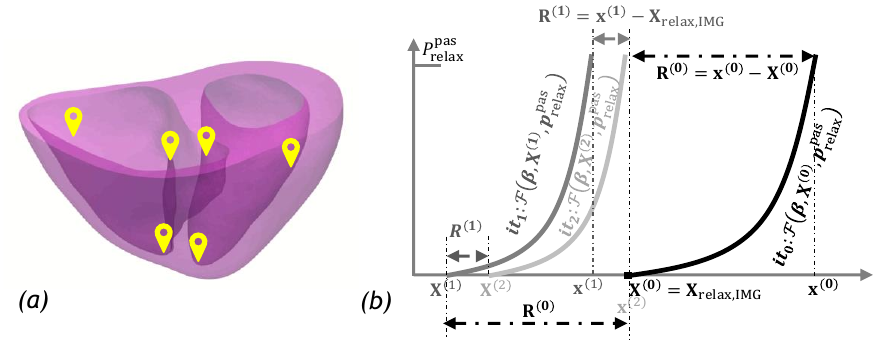}
\caption{(a) Selected landmarks on the biventricle geometry, among which three are selected from the left ventricle and three from the right ventricle. (b) Sellier’s method to determine the stress-free reference configuration by iteratively subtracting the difference in nodal displacements between the FEA-predicted deformed configuration and the image-based myocardium model at its relaxed state from the reference configuration. The deformed configuration is obtained by loading the myocardium with the pressure at the relaxed state using a forward FEA simulation. A relaxed state is defined, for convenience, as the diastasis phase ($\sim$70\% R-R) for the biventricle or the mid-systolic phase ($\sim$20\% R-R) for LA.}
\label{fig_ifea_selliers}
\end{figure}

During each outer iteration $i$ of Algorithm 1 in Fig.~\ref{fig_ifea_algo}, we first conduct a search for the stress-free configuration ($\mathbf{X}^{(i)}$) using Sellier's method (Algorithm 2 in Fig.~\ref{fig_ifea_algo}), with the material parameter set, $\pmb{\beta}^{(i)}$, the relaxed state, $\mathbf{X}_{\text{relax,IMG}}$, and the corresponding pressure, $p^{\text{pas}}_{\text{relax}}$, as input arguments. Details of this search will be discussed later. We then solve a forward FEA problem, $\pmb{\eta}_{\mathrm{FEA}}^{(i)} = \mathcal{F}(\pmb{\beta}^{(i)}, \mathbf{X}^{(i)}, p^{\text{pas}}_T)$, where $p^{\text{pas}}_T$ is the terminal pressure load during passive myocardial expansion, incremented linearly from $0$. The FEA-predicted cavity volumes and myocardial displacements at predefined landmarks (Fig.~\ref{fig_ifea_selliers}a) are gathered at different time points that match image acquisition, into a solution vector $\pmb{\eta}_{\mathrm{FEA}}^{(i)}$. We then construct an objective function based on differences in the cavity volumes (LV and RV, or LA), and the differences in nodal displacements, between FEA-predictions ($\pmb{\eta}_{\text{FEA}}^{(i)}$) and the corresponding data extracted from image segmentations ($\pmb{\eta}_{\text{IMG}}$, Section \ref{ssec:img_seg_reg}) as,

\begin{equation}
    \begin{split}
        E(\pmb{\eta}_{\mathrm{FEA}}^{(i)},\pmb{\eta}_{\mathrm{IMG}}) & = \sum_{J=\{\text{LV},\text{RV}\}/\{\text{LA}\}} E_{J} + \frac{\lambda_{LM}}{n_{LM}} E_{\text{LM}} \\ 
        E_{J} & = \sum_{n=1}^{n_T} \left\vert  \frac{V^n_{\text{FEA},J} - V^n_{\text{IMG},J}}{V^n_{\text{IMG},J}} \right\vert \\
        E_{\text{LM}} & = \sum_{k=1}^{n_{LM}} \sum_{n=1}^{n_T} \left\vert \frac{\mathbf{d}^n_{\text{FEA},k} - \mathbf{d}^n_{\text{IMG},k}} {\mathbf{d}^n_{\text{IMG},k}} \right\vert \\
    \end{split}
    \label{eqn_obj_fun}
\end{equation}

\noindent where $n_{LM}$ is the number of landmarks, $n_T$ denotes the number of image acquisition phases during myocardial passive expansion, $V$ is the cavity volume, and $\mathbf{d}$ is the displacement vector of a landmark with respect to the relaxed configuration. $\lambda_{LM}$ denotes a weighting parameter that adjusts the contribution to the total error from the differences in displacements at the landmarks. We chose $\lambda_{LM}=1$ for all the simulations. The objective function, $\epsilon^{(i)}= E(\pmb{\eta}_{\mathrm{FEA}}^{(i)}, \pmb{\eta}_{\mathrm{IMG}})$, expressed in Eq. (\ref{eqn_obj_fun}), is then minimized using an optimization algorithm, followed by an update of the material parameters, $\pmb{\beta}^{(i+1)} = \mathcal{F}_{\text{opt}}(\epsilon^{(i)},\pmb{\beta}^{(i)})$, depending on the specific optimization method (Section \ref{ssec:outer_optim}). The outer iterations are repeated until either the objective function ($\epsilon^{(i)}$) remains unchanged up to a tolerance parameter (TOL) for at least four iterations or when the iterations reach a maximum limit ($\mathrm{it}_{\mathrm{max}}^{\mathrm{optim}}$). We chose $\mathrm{TOL}=0.1\%$ in this study. Once converged, the algorithm yields the stress-free configuration $\mathbf{X}_{\text{optim}}$ and the best-fit material parameters $\pmb{\beta}_{\mathrm{\text{optim}}}$.

We will now describe Sellier's method used to determine the reference configuration as an inner iteration in our nested optimization framework (Fig.~\ref{fig_ifea_selliers}b).\cite{marx2022robust,sellier2011iterative} The inner algorithm is always initialized with the myocardial configuration at the relaxed state ($\mathbf{X}_{\text{relax,IMG}}$) as the initial reference configuration ($\mathbf{X}^{(0)}$). Likewise, the fiber and sheet orientations are initialized to the ones created using the LDRB method in the relaxed configuration, i.e., $\mathbf{f}^{(0)} = \mathbf{f}_{0_\text{relax,IMG}}, \ \mathbf{s}^{(0)} = \mathbf{s}_{0_\text{relax,IMG}}$. During each inner iteration $j$, we solve a forward FEA problem, denoted as $\mathcal{F}(\pmb{\beta}^{(i)}, \mathbf{X}^{(j)}, p^{\text{pas}}_{\text{relax}})$, where $p^{\text{pas}}_{\text{relax}}$ is the pressure load at the relaxed state of the myocardium, incremented linearly from $0$. The material parameter set, $\pmb{\beta}^{(i)}$, is defined by the outer loop that remains unchanged during these inner iterations. A displacement residual, $\mathbf{R}^{(j)}_{\mathrm{disp}}$, is computed as the difference in nodal displacements between the FEA-predicted configuration and the image-based model at the relaxed state of the myocardium ($\mathbf{R}^{(j)}_{\mathrm{disp}} = \mathbf{x}^{(j)} - \mathbf{X}_{\text{relax,IMG}}$). Finally, an updated reference configuration is obtained by subtracting the displacement residual from the reference configuration at the beginning of this iteration as, $\mathbf{X}^{(j+1)} = \mathbf{X}^{(j)} - \mathbf{R}^{(j)}_{\mathrm{disp}}$. Meanwhile, the deformed fiber directions, $\mathbf{f}^{(j+1)} = \mathbf{F}^{(j)}\mathbf{f}^{(j)}$, are mapped to the updated reference configuration, $\mathbf{X}^{(j+1)}$, as the new set of fiber directions. Here, $\mathbf{F}^{(j)}$ is the deformation gradient tensor for the mapping between the reference configuration $\mathbf{X}^{(j)}$ and the corresponding deformed configuration $\mathbf{x}^{(j)}$. The process is repeated until the displacement residual ($\mathbf{R}^{(j)}_{\mathrm{disp}}$) is lower than a threshold in a suitable norm or the maximum iteration limit $\mathrm{it}_{\mathrm{max}}^{\mathrm{SF}}$ is reached. In practice, we observe that the method converges in 3-4 iterations.

\subsubsection{Specific optimization methods for outer iterations}\label{ssec:outer_optim}

In the previous section, we showed that the objective function (Eq. (\ref{eqn_obj_fun})) is minimized using an optimization method to update the material parameters before proceeding to the next iteration. Here, we provide more details on the various optimization methods employed in the current study, including a custom genetic algorithm (GA), a Bayesian-based optimization method (BO), and the gradient-based Levenberg-Marquardt (LM) algorithm, and analyze their performance in Section~\ref{ssec:comp_optim}.

\paragraph{Genetic algorithm}

The genetic algorithm (GA) is initialized by randomly generating $n_p$ individuals, denoted as $\pmb{\beta}^{(0)}_{1,...,n_p}$, where $n_p$ is the population size, and each $\pmb{\beta}^{(0)}_j$ represents a specific set of material parameters. The parameters are then updated via crossover, mutation, and selection operations, constrained within a predefined set of parameter bounds (Table~\ref{table_param_bounds}). The optimization continues until the FEA-predicted deformation closely matches the motion derived from the image data.

In each outer iteration of the genetic algorithm (Algorithm~1 in Fig.~\ref{fig_ifea_algo}), we generate $\displaystyle n_c = \binom{n_p}{2}$ individuals of “children generation”, $\pmb{\beta}^{(i)}_{1,...,n_c}$, from the individuals of the parent generation, $\pmb{\beta}^{(i-1)}_{1,...,n_p}$, via a crossover operation, 

\begin{equation}
    \begin{cases}
      P(\pmb{\beta}^{(i)}_{j,\text{cross}} = \pmb{\beta}^{(i-1)}_{m}) = \displaystyle 1/3 \\
      P(\pmb{\beta}^{(i)}_{j,\text{cross}} = \pmb{\beta}^{(i-1)}_{n}) = \displaystyle 1/3 \\
      P\Big(\pmb{\beta}^{(i)}_{j,\text{cross}} = \displaystyle\frac{1}{2}(\pmb{\beta}^{(i-1)}_{m} + \pmb{\beta}^{(i-1)}_{n})\Big) = 1/3 
    \end{cases}  
    \label{eqn_ga_crossover}
\end{equation}

\noindent where $P$ is the probability function, and $\displaystyle j=1,...,n_c$ here corresponds to a specific combination of $m=\{1,...,n_p\}$ and $n=\{1,...,n_p\}$ such that $m \neq n$. The mutation operation is then conducted on each parameter set, ${\pmb{\beta}^{(i)}_{j,\text{cross}}}$, as,

\begin{equation}        
    \begin{cases}
      P\big(\pmb{\beta}^{(i)}_{j,\text{cross,mut}} \sim U (\pmb{\beta}_{\mathrm{lb}} , \pmb{\beta}_{\mathrm{ub}}) \big) = \mu_{\mathrm{GA}}\\
      P(\pmb{\beta}^{(i)}_{j,\text{cross,mut}} = \pmb{\beta}^{(i)}_{j,\text{cross}}) = 1 - \mu_{\mathrm{GA}}
    \end{cases}  
    \label{eqn_mutations}
\end{equation} 

\noindent where, $\pmb{\beta}_{\mathrm{lb}}, \pmb{\beta}_{\mathrm{ub}}$ are the set of lower and upper bounds for each parameter in $\pmb{\beta}$, $\mu_{\text{GA}}$ is the mutation rate, and $U(\pmb{\beta}_{\mathrm{lb}}, \pmb{\beta}_{\mathrm{ub}})$ represents a continuous uniform distribution between $\pmb{\beta}_{\mathrm{lb}}$ and $\pmb{\beta}_{\mathrm{ub}}$.  

Therefore, for each parameter sample in the $i^{\text{th}}$ iteration, ${\pmb{\beta}^{(i)}_{j,\text{cross,mut}}}$, we perform Sellier's iterations (Algorithm 2 in Fig.~\ref{fig_ifea_algo}) to estimate the corresponding stress-free reference configuration, $\mathbf{X}^{(i)}_{j,\text{cross,mut}} = \mathcal{F}_{\text{SF}}(\pmb{\beta}^{(i)}_{j,\text{cross,mut}}, \mathbf{X}_{\text{relax,IMG}}, p^{\text{pas}}_{\text{relax}})$, followed by a forward FEA to get the FEA-predicted quantities, $\pmb{\eta}^{(i)}_{j,\text{cross,mut}} = \mathcal{F}(\pmb{\beta}^{(i)}_{j,\text{cross,mut}}, \mathbf{X}^{(i)}_{j,\text{cross,mut}}, p^{\text{pas}}_T)$, and evaluate the objective function, $\epsilon^{(i)}_{j,\text{cross,mut}} =  E(\pmb{\eta}^{(i)}_{j,\text{cross,mut}},\pmb{\eta}_{\mathrm{IMG}})$. The individuals $\pmb{\beta}^{(i)}_j$ of the $i^{\text{th}}$ iteration, are finally obtained from $(\pmb{\beta}^{(i-1)}_{1,...,n_p}, \pmb{\beta}^{(i)}_{j,\text{cross,mut}})$ using a roulette wheel method. The optimization is continued until the convergence criterion defined in Section~\ref{ssec:ifea} is satisfied.

\paragraph{Bayesian optmization}

Bayesian optimization (BO) is a global optimization method that employs a \emph{prior} probability distribution to update the \emph{posterior} distribution over an objective function. The method iteratively explores and exploits the parameter search space to identify the optimal solution. In recent years, there has been growing interest in using Bayesian optimization and its variants for uncertainty quantification to estimate confidence intervals on the predicted material parameters.\cite{torun2022multicomponent,thomas2021kinematics,ryan2022bayesian,ross2023bayesian}

The method relies on a probabilistic surrogate model, often based on Gaussian processes, to approximate the objective function $f (\pmb{\beta})$ ($f$ here is a generic function unrelated to the notation used in Fig.~\ref{fig_ifea_algo}). The available observations $\mathcal{D}_{1:N}$ are typically of the form $({\pmb{\beta}_n,y_n})_{n=1}^N$, where $y_n\thicksim \mathcal{N}(f(\pmb{\beta}_n), \nu)$, with $N$ the total number of observations and $\nu$ denoting the variance of noise introduced into the observations.\cite{frazier2018tutorial} This surrogate model estimates the objective function and its uncertainty across the entire search space. Interested readers are referred to Mackay et al.\cite{mackay1998introduction} for more information about Gaussian processes. 

To determine the optimal parameter set $\pmb{\beta} \in \mathcal{B} \subseteq \mathbb{R}^n$, Bayesian optimization initializes through random explorations within the search space. This initial exploration, also known as random exploration or random sampling, helps gather information about multiple locations in the search space to form initial observations $\mathcal{D}_{n_{ip}}$, where $n_{ip}$ is the number of initial explorations or initial points. Subsequently, the algorithm iteratively explores and exploits the search space to find the optimal solution.

At each outer iteration $i$, an acquisition function $\pmb{\alpha}$ guides the selection of the next location to be evaluated. The acquisition function balances exploration (sampling in regions of high uncertainty) and exploitation (sampling in areas likely to yield better results). Popular acquisition functions include Probability of Improvement (PI), Expected Improvement (EI), and Gaussian Process Upper Confidence Bound (UCB).\cite{frazier2018tutorial} For instance, in the commonly used EI method, the acquisition function is calculated as the expected improvement over the current best, as

\begin{equation}
    \pmb{\alpha}_{\mathrm{EI}}(\pmb{\beta} | \mathcal{D}_{1:N}) = \max \Big\{ \big[\mu(\pmb{\beta}) - f(\pmb{\beta}^{+}) - \xi\big] \Phi (\gamma) + \sigma(\pmb{\beta}) \phi(\gamma), 0 \Big\}
    \label{equ_bo_b}
\end{equation} 

\noindent where $\displaystyle \gamma(\pmb{\beta})=\max \Big\{ \frac{\mu(\pmb{\beta}) - f(\pmb{\beta}^{+}) - \xi}{\sigma(\pmb{\beta})} , 0  \Big\}$. $\mu(\pmb{\beta})$ and $\sigma(\pmb{\beta})$ are mean and standard deviation of $\pmb{\beta}$. $\pmb{\beta}^{+}$ is the location to produce the best objective function value so far. $\Phi$ and $\phi$
 are the Probability Density Function (PDF) and Cumulative Distribution Function (CDF), respectively. $\xi$ determines the amount of exploration during optimization, with higher $\xi$ values leading to more exploration. $\xi$ is set as $0.01$ in this paper. The location to evaluate at the iteration $i$ is determined by 

\begin{equation}
    \pmb{\beta}^{(i)} = \mathrm{argmax}_{\pmb{\beta}}  \ \pmb{\alpha}_{\mathrm{EI }} \big(\pmb{\beta} | \mathcal{D}_{1:N+1}  \big)
\end{equation}

After estimating the objective function at the selected location determined by the acquisition function, the new location and its corresponding objective function value are added to the pool of available observations $\mathcal{D}_{1:N+1} = \mathcal{D}_{1:N}, (\pmb{\beta}^{(i)},y^{(i)})$. This updated information is then used to refine the surrogate model. The process continues iteratively until a convergence criterion (see Sec.~\ref{ssec:ifea}) is met.

In this study, we employed the openly available pragmatic Gaussian-process-based Bayesian optimization framework\footnote{\url{https://github.com/fmfn/BayesianOptimization}}.\cite{snoek2012practical} The hyperparameters for this method include the number of initial points ($n_{ip}$) and the number of maximum iterations ($\mathrm{it}_{\mathrm{max}}^{\mathrm{optim}}$), where the initial points correspond to the steps of random exploration, while the maximum iterations signify the upper limit for the Bayesian optimization process (Table~\ref{table_compare_optim_meth}).

\paragraph{Levenberg-Marquardt algorithm}

The Levenberg-Marquardt (LM) algorithm holds popularity in the field of inverse analysis and optimization owing to its conceptual simplicity and widespread availability in various programming languages.\cite{kauer2002inverse,jiang2021levenberg,fu2021application} To determine the optimal parameter set $\pmb{\beta}$ that minimizes the objective function $f (\pmb{\beta})$ ($f$ here is a generic function unrelated to the notation used in Fig.~\ref{fig_ifea_algo}), the LM algorithm employs an iterative approach. Starting with a user-provided initial parameter set $\pmb{\beta}_0$, the algorithm updates the solution iteratively based on the previous solution and its gradient until the termination criteria are met. This is mathematically expressed as,

\begin{equation}
    \pmb{\beta}^{(i+1)} = \pmb{\beta}^{(i)} - (\nabla^2 f + \mu \mathbf{I})^{-1} \nabla f (\pmb{\beta})
    \label{eqn_lm_b}
\end{equation}

\noindent where $\mu$ is the damping factor and $\mathbf{I}$ is the identity tensor. The LM algorithm exhibits characteristics of the Newton-Raphson method when $\mu=0$, and as $\mu \rightarrow \infty$, it mimics the gradient descent method. Initially set to a small value, the value of $\mu$ is adaptively adjusted based on the step size of the preceding iteration. Specifically, $\mu$ undergoes an increase if the previous step-size is larger and a decrease if smaller. Interested readers are referred to Mor$\mathrm{\acute{e}}$~\cite{more1974levenberg} for a more detailed explanation of the specific method for adjusting $\mu$. This adaptive strategy harnesses the advantages of both methods by (a) ensuring the positive definiteness of $(\nabla^2 f + \mu\mathbf{I})$ by maintaining $\mu>0$, leading to an update along the steepest descent, (b) employing a larger $\mu$ when the solution is distant from the optimal point, facilitating faster convergence, and, (c) employing a smaller $\mu$ when the solution is closer to the optimal point, allowing a Newton-Raphson-like approach resulting in quadratic convergence. In this study, we utilized the built-in LM function in the Python-based scipy library\footnote{\url{https://docs.scipy.org/doc/scipy/reference/generated/scipy.optimize.least_squares.html}}\cite{more1974levenberg} for optimization and comparison with other methods. The initial values are listed in Table~\ref{table_compare_optim_meth}. The convergence criteria are described in \ref{ssec:ifea}.

%%%%%%%%%%%%%%%%%%%%%%%%%%%%%%%%%%%%%%%%%%%%%%%%%%%%%%%%%%%%%%%%%%%%%%%%%%%
%----------------------------   RESULTS   ---------------------------------
%%%%%%%%%%%%%%%%%%%%%%%%%%%%%%%%%%%%%%%%%%%%%%%%%%%%%%%%%%%%%%%%%%%%%%%%%%%

\section{Results}\label{sec:results}

\begin{table}[ht]
\begin{center}
% \begin{minipage}{174pt}
\caption{Cavity volumes (mL) derived from patients' images at different cardiac phases.}
\label{table_cavity_vol_img}
\footnotesize
\begin{tabular}{c c c c c c c}
\toprule
\multirow{3}{*}{Normal}  &   & 70\% R-R & 80\% R-R  & 90\% R-R & 99\% R-R  \\
& LV  & 66.8 & 74.9 & 87.5 & 89.5 \\
& RV & 81.9 & 98.3 & 115.1 & 117.5 \\ 
%\midrule
\arrayrulecolor{black!30}\cmidrule(l{.5em}){3-6}
\multirow{2}{*}{} &   & 20\% R-R & 30\% R-R  & 40\% R-R  \\
& LA & 62.7 & 68.4 & 76.0 \\ 
\arrayrulecolor{black!30}\midrule
\multirow{3}{*}{HOCM1} &   & 72\% R-R & 87\% R-R  & 99\% R-R  \\
& LV & 116.7 & 134.7 & 153.4 \\
& RV & 80.5 & 91.2 & 102.7 \\
\arrayrulecolor{black!30}\midrule
\multirow{3}{*}{HOCM2} &   & 72\% R-R & 82\% R-R  & 95\% R-R  \\
 & LV & 166.0 & 184.8 & 215.7 \\
& RV & 133.0 & 174.5 & 194.1 \\
\arrayrulecolor{black!30}\midrule
\multirow{3}{*}{HOCM3} &   & 70\% R-R & 86\% R-R  & 99\% R-R  \\
 & LV & 122.4 & 134.7 & 141.6 \\
& RV & 86.1 & 100.4 & 113.9 \\
\arrayrulecolor{black}\bottomrule
\end{tabular}
% \end{minipage}
\end{center}
\end{table}

\subsection{Myocardial segmentation and motion analysis}\label{ssec:res_myo_seg}

Table~\ref{table_cavity_vol_img} shows the patients' biventricular cavity volumes during passive expansion, extracted using segmentation methods described in Section~\ref{ssec:img_seg_reg}. Also included are the LA cavity volumes during ventricular systole for the normal subject. An illustrative result of the myocardial motion extracted from the normal subject's images is shown in Fig.~\ref{fig_image_registration}. The basal plane remains relatively flat throughout the deformation but moves longitudinally along the apico-basal direction, while the ventricular apex exhibits minimal movement. When the LA expands passively, the mitral valve plane moves longitudinally toward the ventricular apex, synchronizing with the ventricle, whereas the pulmonary veins remain relatively still.

\begin{table}[ht]
\begin{center}

% \begin{minipage}{174pt}
\footnotesize
\caption{Optimized model parameters for the biventricular and LA myocardium.}
\label{table_optim_ifea_params}
\vspace{4pt}
\begin{tabular}{c c c c c c c}
\toprule
 & $a^{\mathrm{HO}}$ & $b^{\mathrm{HO}}$ & $a_f^{\mathrm{HO}}$  & $b_f^{\mathrm{HO}}$ & $a_s^{\mathrm{HO}}$ & $b_s^{\mathrm{HO}}$ \\
\arrayrulecolor{black!30}\midrule
Control      & $ 1.58\times 10^3 $  & 5.02  & $1.03\times 10^5 $ & 15.1 & $1.48\times 10^4 $ & 10.1 \\
HOCM 1       & $ 3.35\times 10^3 $  & 5.33  & $1.85\times 10^5 $ & 18.5 & $9.48\times 10^4 $ & 11.2 \\
HOCM 2       & $ 4.05\times 10^3 $  & 6.32  & $1.95\times 10^5 $ & 18.0 & $5.48\times 10^4 $ & 9.35 \\
HOCM 3       & $ 6.55\times 10^3 $  & 7.21  & $4.85\times 10^5 $ & 20.1 & $8.48\times 10^4 $ & 11.5 \\
Control (LA) & $ 1.39\times 10^3 $  & 4.98  & $9.47\times 10^4 $ & 15.5 & $1.29\times 10^4 $ & 11.2 \\
\arrayrulecolor{black}\bottomrule
\end{tabular}
\begin{tabular}{c c c c c}
  & $C^{\mathrm{GM}}$   & $b_f^{\mathrm{GM}}$ & $b_s^{\mathrm{GM}}$  & $b_{fs}^{\mathrm{GM}}$ \\
\arrayrulecolor{black!30}\midrule
Control   & $331 $  & 281  & 2.52 & 18.9 \\     
HOCM 1    & $639 $  & 312  & 4.27 & 23.7   \\   
HOCM 2    & $485 $  & 301  & 8.58 & 25.8    \\   
HOCM 3    & $1410 $  & 351  & 9.83 & 32.5   \\   
\arrayrulecolor{black}\bottomrule
\end{tabular}
% \end{minipage}
\end{center}
\begin{tablenotes}
    \centering
    \vspace{-3pt}
    \footnotesize
    %\scriptsize
    \item $a^{\mathrm{HO}}_{(\cdot)}$: $\text{dyn}/\text{cm}^2$; $C^{\mathrm{GM}}$: $\text{dyn}/\text{cm}^2$. $b_{(\cdot)}$ is dimensionless.
\end{tablenotes}
\end{table}

\subsection{Optimized constitutive model parameters using inverse FEA} \label{ssec:res_optim_params}

We characterize the mechanics of patient-specific ventricles during passive expansion using inverse modeling and optimization described in Section (\ref{ssec:ifea}). Each optimization involves tuning six material parameters for the HO model ($a^\mathrm{HO}, a_f^\mathrm{HO}, a_s^\mathrm{HO}, b^\mathrm{HO}, b_f^\mathrm{HO}, b_s^\mathrm{HO}$) and four parameters for the GM constitutive model ($C^\mathrm{GM}, b_f^\mathrm{GM}, b_s^\mathrm{GM}, b_{fs}^\mathrm{GM}$). For all the subjects considered in the study, the GA-based optimization converges in approximately 7 hours on an Intel Core i9-13900 processor (24 cores) with 128 GB of memory, employing a mesh with $\sim$3mm resolution and $\sim$90,000 tetrahedral elements. A comparison with other optimization methods and their sensitivity to initial conditions is performed in Section~\ref{ssec:comp_optim}.

Table~\ref{table_optim_ifea_params} summarizes the GA-based optimal constitutive model parameters for the biventricular and left atrial myocardium for all the subjects considered in this study. The parameter set $\{a^\mathrm{HO}, a_f^\mathrm{HO}, a_s^\mathrm{HO}\}$ denotes stiffness-like coefficients while $\{b^\mathrm{HO}, b_f^\mathrm{HO}, b_s^\mathrm{HO}\}$ govern the exponentials for the HO model (Eq. \ref{eq_const_ho}). Likewise, $C^\mathrm{GM}$ represents stiffness-like parameter while $\{b_f^\mathrm{GM}, b_s^\mathrm{GM}, b_{fs}^\mathrm{GM}\}$ comprise exponential coefficients for the GM model (Eq. \ref{eq_const_gm}). 

Generally, the HO stiffness-like parameters $a^\mathrm{HO}, a_f^\mathrm{HO}, \text{and } a_s^\mathrm{HO}$ are of the order of $10^3, 10^5, \text{and } 10^4\text{ dyn}\cdot\text{cm}^{-2}$, respectively for all the patients in this study, whereas the HO exponential parameters span $5-20$. On the otherhand, the GM stiffness-like parameter $C^\mathrm{GM}$ is around $10^2 - 10^3~\text{dyn}\cdot\text{cm}^{-2}$ for both normal and HOCM patients while the exponential parameters $b_f^{\mathrm{GM}},b_s^{\mathrm{GM}}, \text{and } b_{fs}^{\mathrm{GM}}$ are in the order of $10^2, 10^0, \text{and } 10^1$, respectively.

Our results suggest that the normal subject’s ventricle exhibits a lower stiffness than that of the HOCM patients. This observation is consistent for both HO and GM constitutive models. We also observed variability in the model parameters between HOCM patients. For instance, the fiber stiffness ($a_f^\mathrm{HO}$) varies from $18.5-48.5 \text{kPa}$ among the three HOCM patients when using the HO model. Likewise, $C^\mathrm{GM}$ varies between $485 \text{ and } 1410 \text{dyn}\cdot\text{cm}^{-2}$ when using the GM constitutive model. Similar observations can be made for all other model parameters as well. %This underscores the importance of performing patient-specific modeling irrespective of the constitutive model employed.

\begin{figure}[ht]
\centering
\includegraphics[width=1\textwidth]{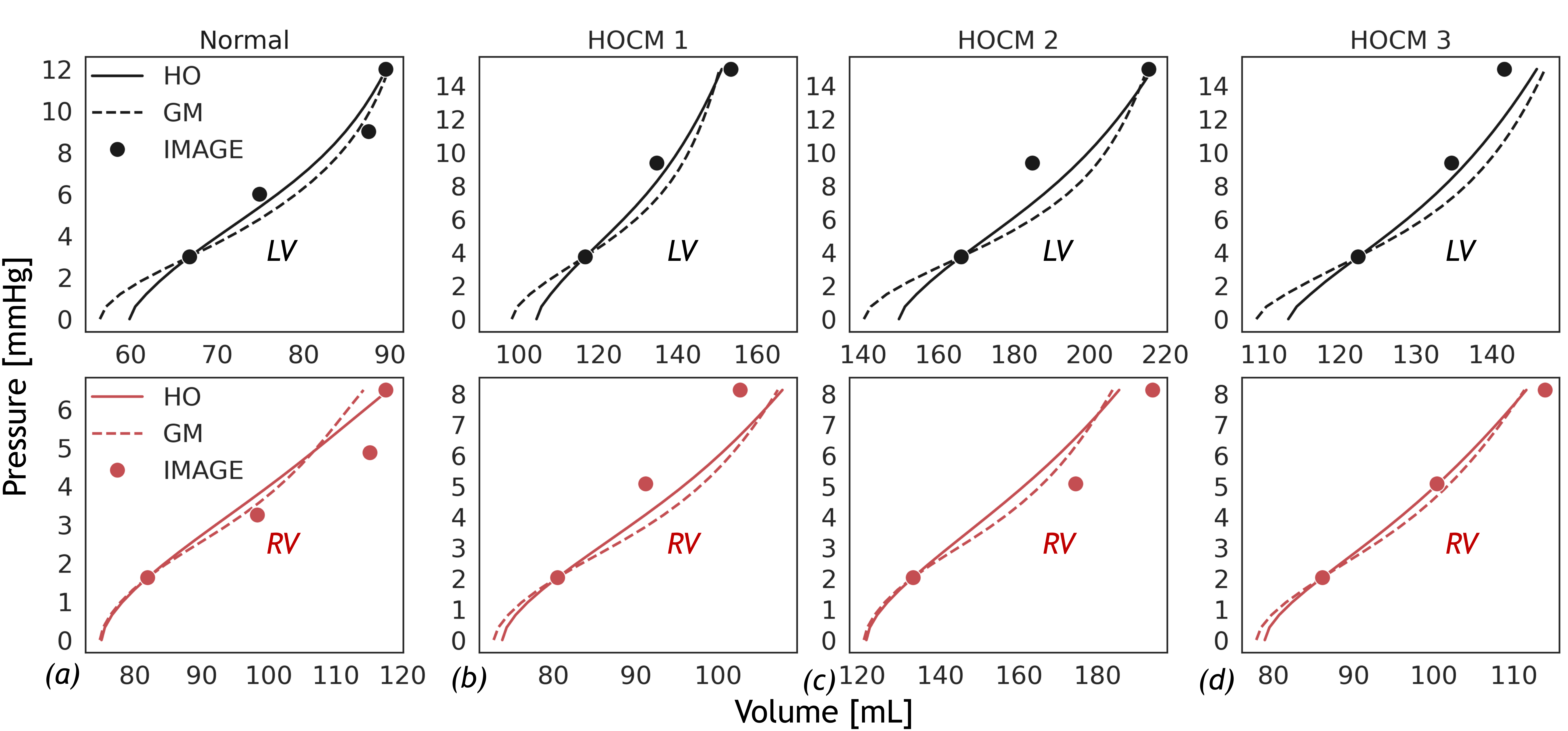}
\caption{Comparison of pressure-volume (P-V) profiles during passive ventricular expansion between inverse FEA and image data for (a) a normal subject and (b-d) the three HOCM patients. LV: left ventricle; RV: right ventricle; HO: Holzapfel-Ogden model; GM: Guccione-McCulloch model;}
\label{fig_pv_profiles_biv}
\end{figure} 

\begin{figure}[ht]
\centering
\includegraphics[width=1.0\textwidth]{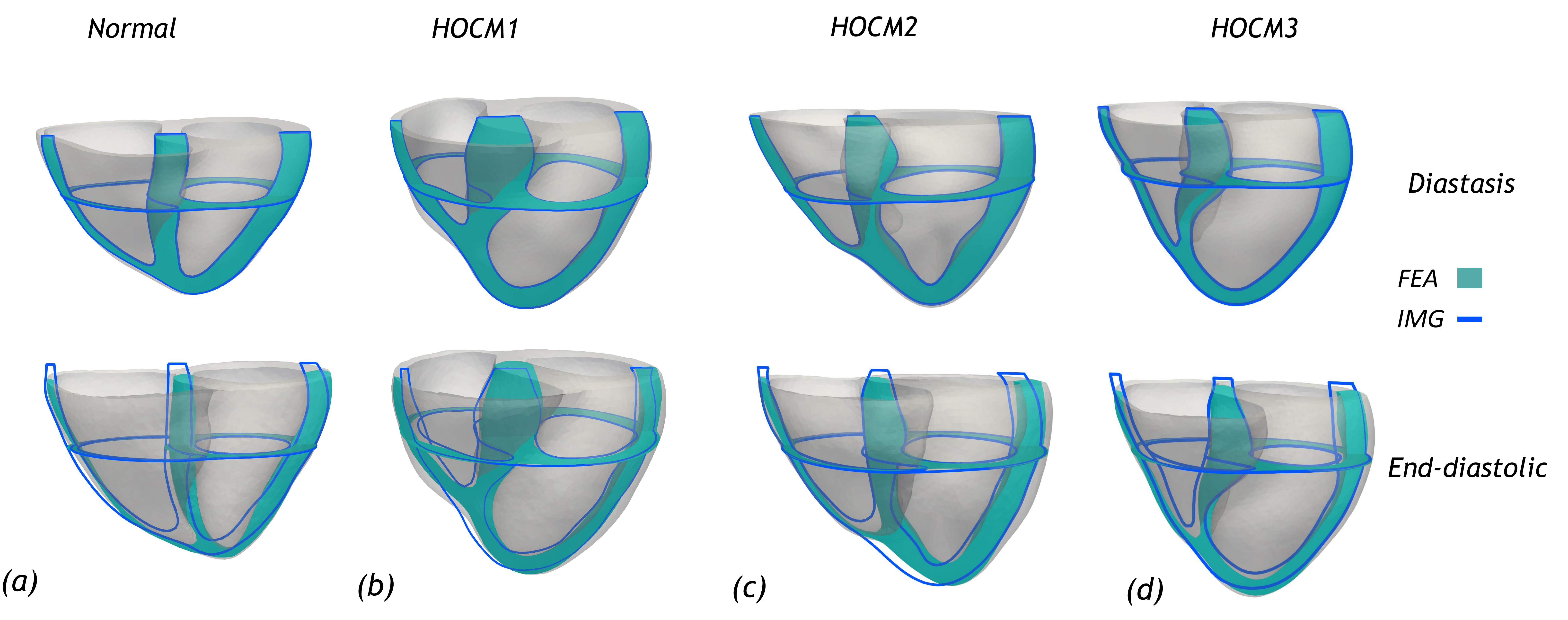}
\caption{Comparison of FEA-predicted deformed configuration against image data along longitudinal and transverse mid-planes of the biventricular myocardium. (a) normal subject; (b-d) HOCM patients; (top) relaxed configuration (mid-diastasis); (bottom) end-diastolic configuration.}
\label{fig_compare_section_biv}
\end{figure} 

\begin{figure}[ht]
\centering
\includegraphics[width=0.9\textwidth]{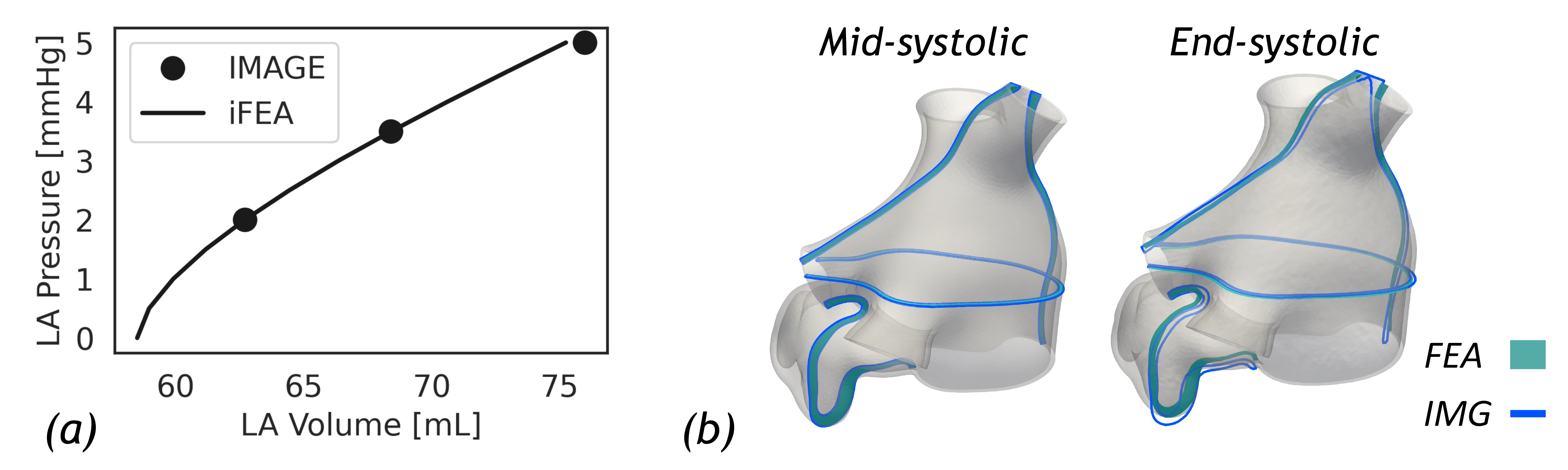}
\caption{Comparison of LA passive mechanics in a healthy subject between FEA and image data. (a) P-V profile. (b) Comparison of FEA-predicted deformation against image data along longitudinal and transverse slices during mid-systole (left) and end-systole (right).}
\label{fig_pv_section_la}
\end{figure}

\subsection{Comparison between FEA-predictions and image data} \label{ssec:res_comp_fea_img}

A comparison of LV and RV pressure-volume (P-V) profiles during passive filling shows a reasonable agreement between simulation predictions and data extracted from patients’ images for the normal subject and the three HOCM patients (Fig.~\ref{fig_pv_profiles_biv}). Marginal differences were noted between the orthotropic HO model and the transversely isotropic GM model for both LV and RV across all the subjects. Likewise, the P-V profile shows excellent agreement for the LA myocardium during passive filling (Fig.~\ref{fig_pv_section_la}a). The sectional views of the FEA-predicted deformed configurations, along longitudinal and transverse planes, align reasonably well with the image data for biventricle and LA myocardium (Fig.~\ref{fig_compare_section_biv},~\ref{fig_pv_section_la}b). 

The relative errors in cavity volumes are reasonably within 10\% for all patients and across all cardiac phases during passive filling (Tables~\ref{table_rel_vol_ho},~\ref{table_rel_vol_gm}). With the HO model, most of the FEA-predicted biventricular cavity volumes are within 5\% of the image-based data for all the patients and at all cardiac phases (Table~\ref{table_rel_vol_ho}). One exception is the RV volume at 90\% R-R, which exhibits a slightly higher deviation (7.4\%) for the normal subject. On the other hand, we note relatively higher differences for the GM model in the FEA-predicted data compared to the images (Table~\ref{table_rel_vol_gm}). The LA myocardium shows remarkable agreement with image data, maintaining the relative errors under 1\% (Table~\ref{table_rel_vol_ho}).

\begin{table}[ht]
\begin{center}
% \begin{minipage}{174pt}
\caption{Percentage relative error in cavity volumes between FEA and image data using the Holzapfel-Ogden (HO) constitutive model. Errors above 5\% are \bf{\textcolor{c3}{highlighted}}.}
\label{table_rel_vol_ho}
\vspace{4pt}
\footnotesize
\begin{tabular}{c c c c c c c}
\toprule
\multirow{3}{*}{Normal}  &   & 70\% R-R & 80\% R-R  & 90\% R-R & 99\% R-R  \\
& LV  & 0.05 & 2.7 & 3.0 & 0.07 \\
& RV & 0.06 & 4.2 & \bf{\textcolor{c3}{7.4}} & 0.4 \\ 
\arrayrulecolor{black!30}\cmidrule(l{.5em}){3-6}
\multirow{2}{*}{} &   & 20\% R-R & 30\% R-R  & 40\% R-R  \\
& LA & 0.01 & 0.05 & 1.0 \\ 
\arrayrulecolor{black!30}\midrule
\multirow{3}{*}{HOCM1} &   & 72\% R-R & 87\% R-R  & 99\% R-R  \\
& LV & 0.05 & 1.8 & 1.5 \\
& RV & 0.02 & 4.4 & \bf{\textcolor{c3}{5.0}} \\
\arrayrulecolor{black!30}\midrule
\multirow{3}{*}{HOCM2} &   & 72\% R-R & 82\% R-R  & 95\% R-R  \\
& LV & 0.05 & 0.6 & 0.5 \\
& RV & 0.02 & 0.9 & 4.4 \\
\arrayrulecolor{black!30}\midrule
\multirow{3}{*}{HOCM3} &   & 70\% R-R & 86\% R-R  & 99\% R-R  \\
& LV & 0.06 & 0.9 & 3.0 \\
& RV & 0.01 & 0.7 & 2.1 \\
\arrayrulecolor{black}\bottomrule
\end{tabular}
% \end{minipage}
\end{center}
\end{table}

\begin{table}[ht]
\begin{center}
% \begin{minipage}{174pt}
\caption{Percentage relative error in cavity volumes between FEA and image data using the Guccione-McCulloch (GM) constitutive model. Errors above 5\% are \bf{\textcolor{c3}{highlighted}}.}
\label{table_rel_vol_gm}
\vspace{4pt}
\footnotesize
\begin{tabular}{c c c c c c c}
\toprule
\multirow{3}{*}{Normal}  &   & 70\% R-R & 80\% R-R  & 90\% R-R & 99\% R-R  \\
 & LV & 0.08 & \bf{\textcolor{c3}{5.6}} & 1.5 & 0.4 \\
 & RV & 0.02 & 2.2 & \bf{\textcolor{c3}{7.4}} & 2.8 \\ 
\arrayrulecolor{black!30}\midrule
\multirow{3}{*}{HOCM1} &   & 72\% R-R & 87\% R-R  & 99\% R-R  \\
 & LV & 0.05 & 4.2 & 1.9 \\
 & RV & 0.02 & \bf{\textcolor{c3}{6.7}} & 4.5 \\
\arrayrulecolor{black!30}\midrule
\multirow{3}{*}{HOCM2} &   & 72\% R-R & 82\% R-R  & 95\% R-R  \\
 & LV & 0.07 & \bf{\textcolor{c3}{8.7}} & 0.2 \\
 & RV & 0.01 & \bf{\textcolor{c3}{5.7}} & \bf{\textcolor{c3}{5.3}} \\
\arrayrulecolor{black!30}\midrule
\multirow{3}{*}{HOCM3} &   & 70\% R-R & 86\% R-R  & 99\% R-R  \\
 & LV & 0.03 & 3.0 & 3.8 \\
 & RV & 0.0  & 0.8 & 2.3 \\
\arrayrulecolor{black}\bottomrule
\end{tabular}
% \end{minipage}
\end{center}
\end{table}

\subsection{Comparison between optimization methods}\label{ssec:comp_optim} 

Here, we compare the performance of the various optimization methods used in the outer iterations as part of the inverse FEA (Section~\ref{ssec:outer_optim}). Without loss of generality, we compare these algorithms for the biventricular model of the healthy subject, modeled using the HO constitutive model. The boundary conditions, loading profiles, solver parameters, and numerics are consistent across all the cases. We compare the optimization methods for computational performance by assessing their convergence and elapsed wall time, as well as their sensitivity to initial conditions. In particular, we vary the number of individuals ($n_p$) for the GA algorithm from 4 to 5. For the BO method, we change the hyperparameter, the number of initial points ($n_{ip}$), from 2 to 4. We chose two sets of initial parameters from the literature for the gradient-based LM method.\cite{palit2017passive,holzapfel2009constitutive} Finally, we compare the P-V profiles against image data for all cases.

\begin{table}[ht]
\begin{center}
% \begin{minipage}{174pt}
\footnotesize
\caption{Performance comparison of the different optimization methods for the outer iterations (control case only).}
\label{table_compare_optim_meth}
\vspace{4pt}
\begin{tabular}{c c c c }
\toprule
Method  & $\epsilon_\infty$  & Comput. Time & Initial Parameters \\
\arrayrulecolor{black!30}\midrule
 GA1    & 2.32  & 6.8 hours  & Individual: 4, Max iter: 30  \\
 GA2    & 2.35  & 8.9 hours  & Individual: 5, Max iter: 30  \\
 BO1    & 2.37  & 7.8 hours  & Initial points: 2, Max iter: 180\cite{snoek2012practical}    \\
 BO2    & 2.34  & 8.1 hours  & Initial points: 4, Max iter: 180\cite{snoek2012practical}    \\
 LM1    & 2.53  & 23.8 hours  & Initial values in~\cite{palit2017passive} \\
 LM2    & 2.93  & 35.5 hours  & Initial values in~\cite{holzapfel2009constitutive} \\
\arrayrulecolor{black}\bottomrule
\end{tabular}
\vspace{4pt}
\begin{tabular}{c c c c c c c}
Method & $a^{\mathrm{HO}}$ & $b^{\mathrm{HO}}$ & $a_f^{\mathrm{HO}}$  & $b_f^{\mathrm{HO}}$ & $a_s^{\mathrm{HO}}$ & $b_s^{\mathrm{HO}}$ \\
\arrayrulecolor{black!30}\midrule
 GA1    & $1.58\times 10^3$  & 5.02  & $1.03\times 10^5$ & 15.1 & $1.48\times 10^4$ & 10.1  \\
 GA2    & $1.47\times 10^3$  & 5.17  & $9.87\times 10^4$ & 15.5 & $1.37\times 10^4$ & 11.2   \\
 BO1    & $1.50\times 10^3$  & 4.98  & $1.13\times 10^5$ & 14.7 & $1.32\times 10^4$ & 11.5    \\
 BO2    & $1.71\times 10^3$  & 5.08  & $1.07\times 10^5$ & 14.5 & $1.36\times 10^4$ & 9.8    \\
 LM1    & $1.43\times 10^3$  & 5.49  & $9.47\times 10^4$ & 16.9 & $1.43\times 10^4$ & 10.2  \\
 LM2    & $2.58\times 10^3$  & 5.85  & $1.07\times 10^6$ & 9.03 & $1.09\times 10^5$ & 6.12  \\
\arrayrulecolor{black}\bottomrule
\end{tabular}
% \end{minipage}
\end{center}
\begin{tablenotes}
\footnotesize
\item $\epsilon_\infty$ is a measure of the objective function at convergence. GA: Genetic Algorithm; BO: Bayesian Optimization; LM: Levenberg-Marquardt. $a^{\mathrm{HO}}_{(\cdot)}$: $\text{dyn}/\text{cm}^2$; $b_{(\cdot)}$ is dimensionless. LM1 parameters are initialized by averaging the data for five patients from Table 2 of Palit et al.~\cite{palit2017passive}. LM2 parameters are extracted from Figure 7 of Holzapfel et al.~\cite{holzapfel2009constitutive}.   
\end{tablenotes}
\end{table}

\begin{figure}[ht]
\centering
\includegraphics[width=0.8\textwidth]{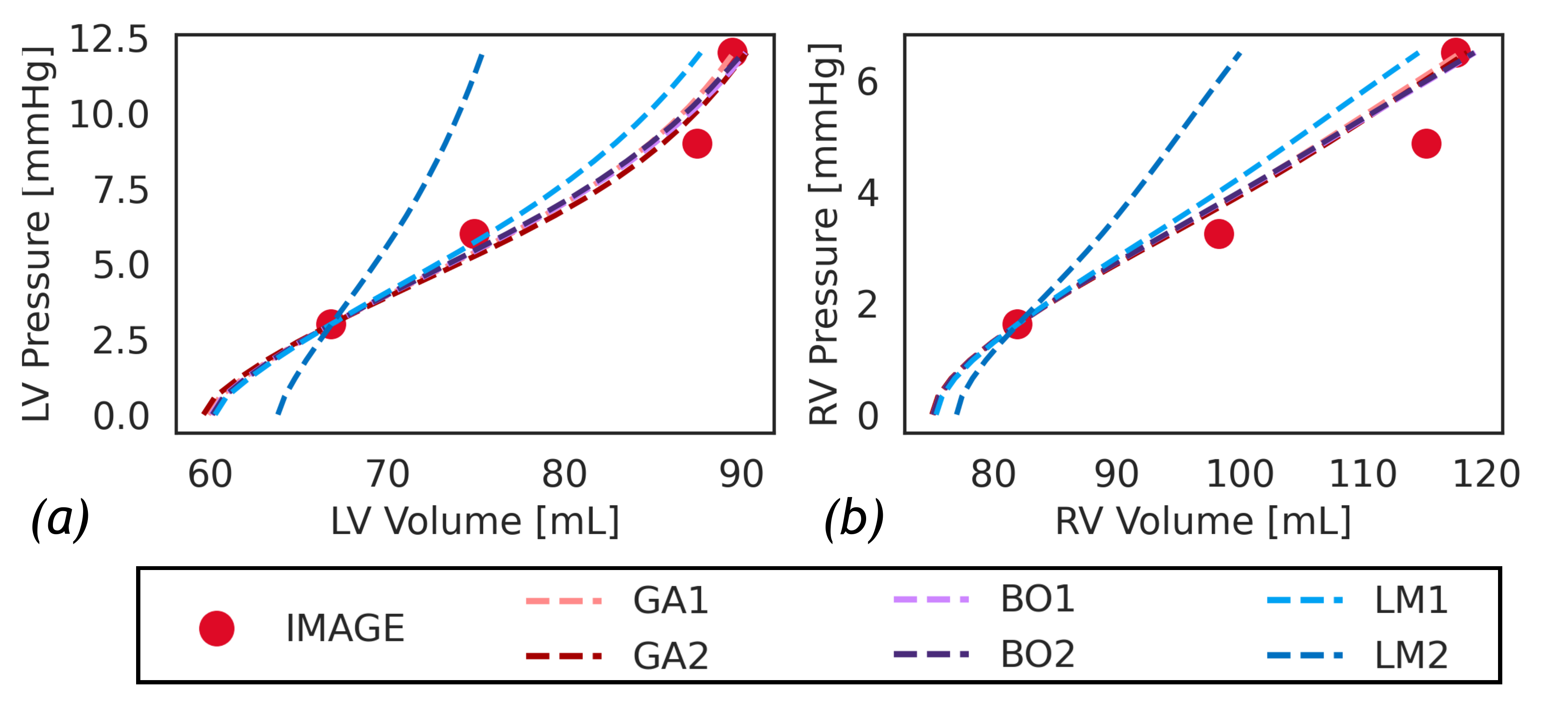}
\caption{Comparison of (a) left and (b) right ventricular P-V profiles for different optimization methods and initial conditions. Solid markers represent image data. GA: Genetic Algorithm; BO: Bayesian Optimization; LM: Levenberg-Marquardt. LV: left ventricle; RV: right ventricle.}
\label{fig_fea_algor}
\end{figure}

We observe that $\epsilon_\infty$, a measure of the objective function after convergence, is similar in magnitude for the GA and BO methods but differs by $\sim$10\% for the LM method (GA1 vs. BO1 vs. LM1 in Table~\ref{table_compare_optim_meth}). At the same time, while BO takes only 15\% more computational time than GA until convergence (6.8h for GA1 vs. 7.8h for BO1, Table~\ref{table_compare_optim_meth}), the LM method is substantially more expensive (3.5x) compared to GA (6.8h for GA1 vs. 23.8h for LM1, Table~\ref{table_compare_optim_meth}). Regarding the constitutive model parameters, the GA and BO methods differ mostly within 10\%, the only exception being $b_s^{\mathrm{HO}}$ parameter, which is higher by $\sim$14\% for the BO method ($b_s^{\mathrm{HO}}$=10.1 for GA1 vs. $b_s^{\mathrm{HO}}$=11.5 for BO1, Table~\ref{table_compare_optim_meth}). Likewise, the LM model parameters also vary within 10\% of that of the GA method, except for $b_f^{\mathrm{HO}}$ that is slightly higher by 12\% (GA1 vs. LM1 in Table~\ref{table_compare_optim_meth}). However, despite these differences in the optimal material parameters from the three methods, we see a close alignment between the GM and BO in the P-V profiles, while the LM shows noticeable deviation (compare GA1, BO1, and LM1 in Fig.~\ref{fig_fea_algor}).

We also note that the LM method is highly sensitive to the initial conditions. By initializing the parameters from different sources in the literature, we find that LM not only takes substantially more computational effort, but leads to an entirely different set of material parameters, some of which are orders of magnitude different (LM1 vs. LM2 in Table~\ref{table_compare_optim_meth}), and a misaligned P-V profile (LM2 in Fig.~\ref{fig_fea_algor}). On the contrary, GA and BO methods exhibit lower sensitivity to input parameters. For instance, changing $n_p$ from 4 to 5 for the GA method results in at most 11\% change in the material parameters, at 30\% increase in computational cost (GA1 vs. GA2 in Table~\ref{table_compare_optim_meth}). Likewise, changing $n_{ip}$ from 2 to 4 results in, at most 15\% change in material parameters at a marginal increase in computational cost (BO1 vs. BO2 in Table~\ref{table_compare_optim_meth}). Nevertheless, the P-V profiles are relatively less affected by changes to input parameters for both these optimization methods (GA vs. BO in Fig.~\ref{fig_fea_algor}).

%%%%%%%%%%%%%%%%%%%%%%%%%%%%%%%%%%%%%%%%%%%%%%%%%%%%%%%%%%%%%%%%%%%%%%%%%%%
%------------------------    DISCUSSION   ---------------------------------
%%%%%%%%%%%%%%%%%%%%%%%%%%%%%%%%%%%%%%%%%%%%%%%%%%%%%%%%%%%%%%%%%%%%%%%%%%%

\section{Discussion}\label{sec:discussion}

The primary objective of this work is to develop a computational framework for efficiently estimating patient-specific myocardial material parameters during passive expansion using time-dependent image data. We achieved this by introducing a nested multi-level optimization scheme that simultaneously estimates passive mechanical parameters of the myocardium and the reference stress-free configuration that best approximate myocardial deformation from the image data (Section~\ref{ssec:ifea}, Fig.~\ref{fig_ifea_algo}). The framework allows flexibility in choosing a traditional optimization method at the outer level that iterates on the material parameters, while the inner level employs a modified Sellier’s algorithm to estimate the stress-free configuration (Fig.~\ref{fig_ifea_algo}). 

Using time-resolved 3D CT data as input, we applied the framework to characterize the passive mechanics of the biventricular myocardium in a healthy subject and three HOCM patients (Table~\ref{table_optim_ifea_params}). We also demonstrated the framework's utility in characterizing the left atrial myocardium in a healthy subject. A qualitative comparison of PV curves and sectional profiles in the deformed configuration showed a reasonable agreement between FEA predictions and image data (Figs.~\ref{fig_pv_profiles_biv},~\ref{fig_compare_section_biv}). A quantitative comparison resulted in less than 5\% error in cavity volumes between FEA and image data using the HO model at most phases of the passive filling (Table~\ref{table_rel_vol_ho}), and less than 10\% error when using the GM constitutive model (Table~\ref{table_rel_vol_gm}). The performance was even more remarkable for the LA myocardium qualitatively (Fig.~\ref{fig_pv_section_la}) and quantitatively with less than 1\% relative error in cavity volumes. This reaffirms the robustness and credibility of the current modeling framework to personalize myocardial mechanics using patient data. At the same time, the variability in the estimated material parameters for the HOCM patients (Table~\ref{table_optim_ifea_params}) underscores the importance of performing patient-specific analysis using inverse modeling, ultimately aiming to develop personalized diagnostic and treatment plans. 

A comparison of different optimization methods in the outer iterations revealed that the identified material parameters are sensitive to the choice of the optimization method and the settings used for initializing these methods (Table~\ref{table_compare_optim_meth}). However, the genetic algorithm and Bayesian optimization methods were robust in capturing image-based deformation and P-V profiles (Fig.~\ref{fig_fea_algor}), with less variability in the estimated optimal material parameters. On the contrary, the gradient-based LM method was computationally expensive and highly sensitive to initial conditions, potentially leading to erroneous results (Fig.~\ref{fig_fea_algor}). 

Although we acknowledge that the manual segmentation approach is tedious, we have continued to adopt this approach in our current work for its accuracy and control. Interestingly, our previously developed registration technique to extract myocardial motion was ineffective in tracking the biventricular geometry in HOCM patients, thereby underpredicting the cavity volumes by nearly 30-70\% during diastole.\cite{vedula2017method} In the future, we will work toward improving HeartDeformNet’s capabilities to automatically segment myocardium in HOCM patients and include additional features such as LAA and papillary muscles in its training dataset.

We designed the objective function of the optimization algorithm to minimize discrepancies in integral measures, such as the cavity volumes, and local nodal displacements at selected landmarks, allowing a finer control of the deformation. These landmarks act as anchor points around which the ventricles deform and, therefore, are chosen at anatomical locations, such as the ventricular free wall, basal plane, septum, and apex. However, adding too many landmarks may not improve agreement with the images; instead, it may have the opposite effect of poor alignment with image data. This is because image-based ventricular motion is obtained by tracking the changes in the intensity field (Eulerian) but does not track a fixed material point on the myocardium (Lagrangian). Consequently, image-based motion does not capture the true deformation of the myocardium, e.g., twisting motion, but captures accurate changes in topology (e.g., cavity volume). On the other hand, the FEA employs a rigorous continuum mechanics framework that models fine-scale deformation, such as fiber deformation, which governs ventricular torsion. Therefore, in the current framework, we only consider three landmarks on the left and right ventricular endocardium.

We will now discuss the sensitivity of the model’s predictions to various underlying modeling assumptions and simplifications.

\begin{figure}[ht]
\centering
\includegraphics[width=0.8\textwidth]{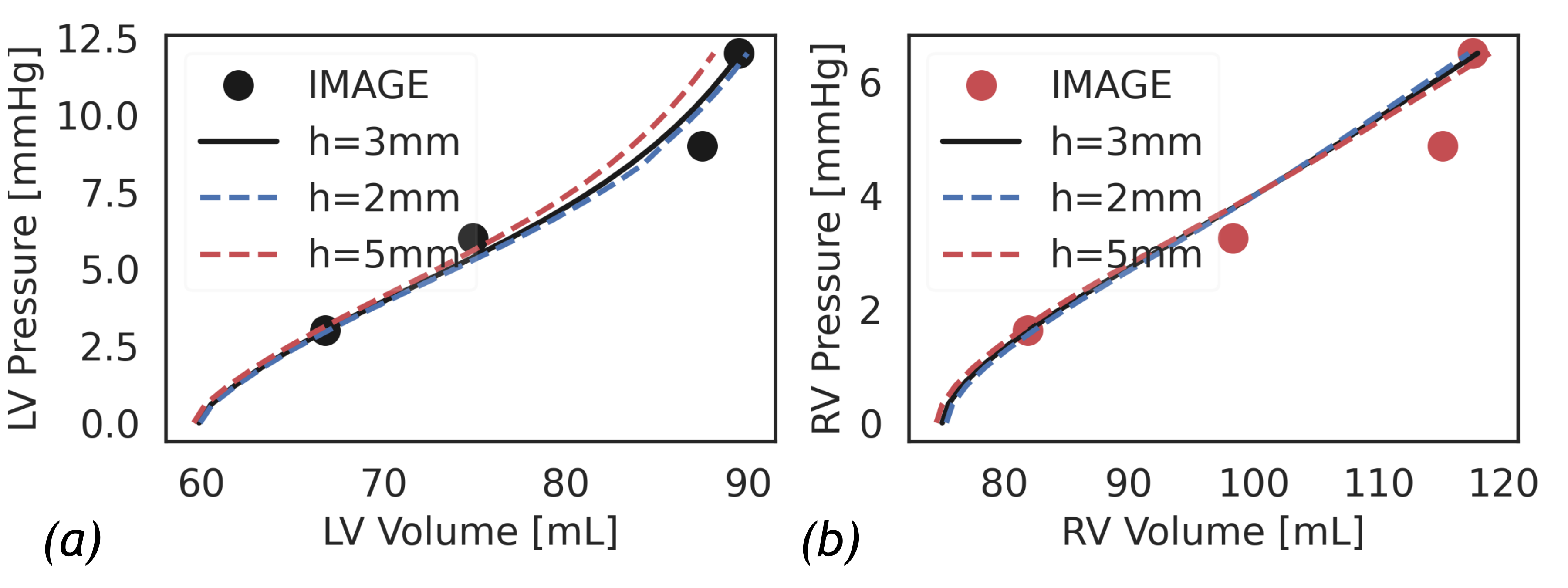}
\caption{Effect of the mesh element size on the (a) left and (b) right ventricular pressure-volume (P-V) curves. The fine (h=2mm) and medium (h=3mm) meshes show nearly overlapping P-V curves, while the coarse mesh exhibits deviation toward end-diastole. LV: left ventricle; RV: right ventricle.}
\label{fig_mesh_size}
\end{figure}

\subsection{Effect of the mesh size}\label{ssec:mesh_size}

Here, we assess the sensitivity of our biventricular mechanics model to the mesh edge size. In particular, we evaluate the impact of element size on the P-V curves of the left and right ventricles (Fig.~\ref{fig_mesh_size}). We compare three different element sizes for the healthy biventricular myocardium: (i) a coarse mesh with h=5mm edge size, (ii) a medium-sized mesh with edge size, h=3mm, and (iii) a fine mesh with edge size, h=2mm, resulting in approximately 15K, 75K, and 250K elements, respectively. We employed the optimal HO material parameters for all the cases (Table~\ref{table_optim_ifea_params}) and applied boundary conditions and solver parameters as described in Section~\ref{ssec:num_proc} and Table~\ref{table_fix_params_biv}. As expected, the total CPU time increases substantially from coarse to fine meshes (coarse: 1.3min, median: 4.7min, fine: 24min) while only minimally affecting the P-V curves. The medium and fine meshes exhibit a nearly overlapping P-V curve, although the coarse mesh results in a noticeable deviation toward end-diastole (Fig.\ref{fig_mesh_size}). Consequently, to balance between accuracy and efficiency, we chose the medium element size for iFEA that involves hundreds of iterations as part of the optimization process (Section~\ref{ssec:ifea}).

\begin{figure}[ht]
\centering
\includegraphics[width=0.8\textwidth]{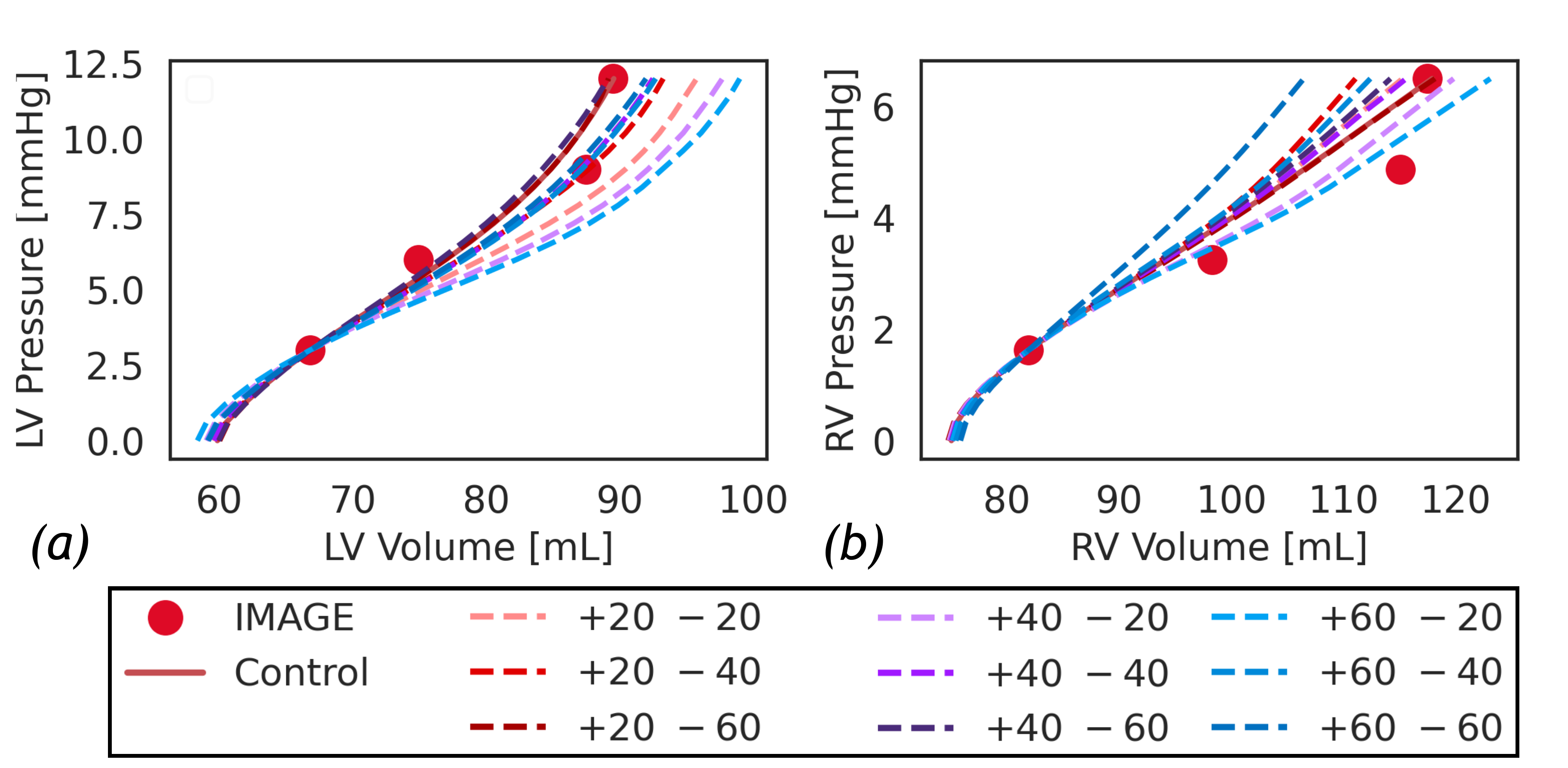}
\caption{Effect of longitudinal fiber angles at the endocardium ($\alpha_\mathrm{endo}$) and epicardium ($\alpha_\mathrm{epi}$) on the (a) left ventricular and (b) right ventricular pressure-volume (P-V) profiles. Solid dots represent data from images while the thick solid line represents the baseline case with $\alpha_\mathrm{endo}$ and $\alpha_\mathrm{epi}$ set to $40^\circ$ and $-50^\circ$, respectively. A positive value indicates $\alpha_\mathrm{endo}$, while a negative value denotes $\alpha_\mathrm{epi}$. The darker the color, the greater the absolute value of the fiber angle. LV: left ventricle; RV: right ventricle.}
\label{fig_fea_orient}
\end{figure}

\subsection{Effect of fiber orientation} \label{ssec:fiber_orient}

In this study, we employed a rule-based method to assign fiber directions in the myocardium, linearly varying along the transmural direction from $40^\circ$ at the endocardium to $-50^\circ$ at the epicardium.\cite{bayer2012novel} However, data from diffusion tensor imaging (DTMRI) in animal models reveal local variations in fiber directions and fiber dispersion.\cite{wang2009modelling,guan2020effect} Other variants of simplified rule-based fiber direction models were analyzed for ventricular pump function and compared against DTMRI-based models.\cite{guan2020effect,avazmohammadi2019contemporary} Different settings for fiber angles ($\alpha_\mathrm{endo}, \alpha_\mathrm{epi}$) and sheet angles ($\beta_\mathrm{endo}, \beta_\mathrm{epi}$) were applied by various groups.\cite{bayer2012novel,pfaller2019importance,guan2020effect,palit2017passive} To assess the impact of variations in these angles on the ventricular passive mechanics, we conducted sensitivity tests by varying the longitudinal fiber angles $\alpha_\mathrm{endo}$ and $\alpha_\mathrm{epi}$, while keeping the transverse sheet orientation $\beta_\mathrm{endo}$ and $\beta_\mathrm{epi}$ fixed. We varied the parameter set $\{\alpha_\mathrm{endo}, \alpha_\mathrm{epi}\}$ between $\{20^\circ, -20^\circ\}$ and $\{60^\circ, -60^\circ\}$ in increments of $20^\circ$ each, comprising a total of 9 cases. For this analysis, we employed the optimal HO constitutive model parameters for the biventricular myocardium (Table~\ref{table_optim_ifea_params}), retained all other simulation parameters as given in Table~\ref{table_fix_params_biv} and followed the numerics described in Section~\ref{ssec:num_proc}. We found that the fiber directions didn’t substantially alter the reference configuration or the deformed state near the early phase of passive expansion. However, as the pressure load increases, we found that the P-V curves are highly sensitive to fiber directions, fanning out from mid-diastasis toward the end of the diastole (Fig.~\ref{fig_fea_orient}), suggesting that the estimated material parameters could be highly sensitive to the fiber orientations predefined using rule-based methods. This issue may addressed in the future with advances in \textit{in vivo} myofiber imaging.~\cite{khalique2020diffusion,guan2020effect}

\begin{figure}[ht]
\centering
\includegraphics[width=0.8\textwidth]{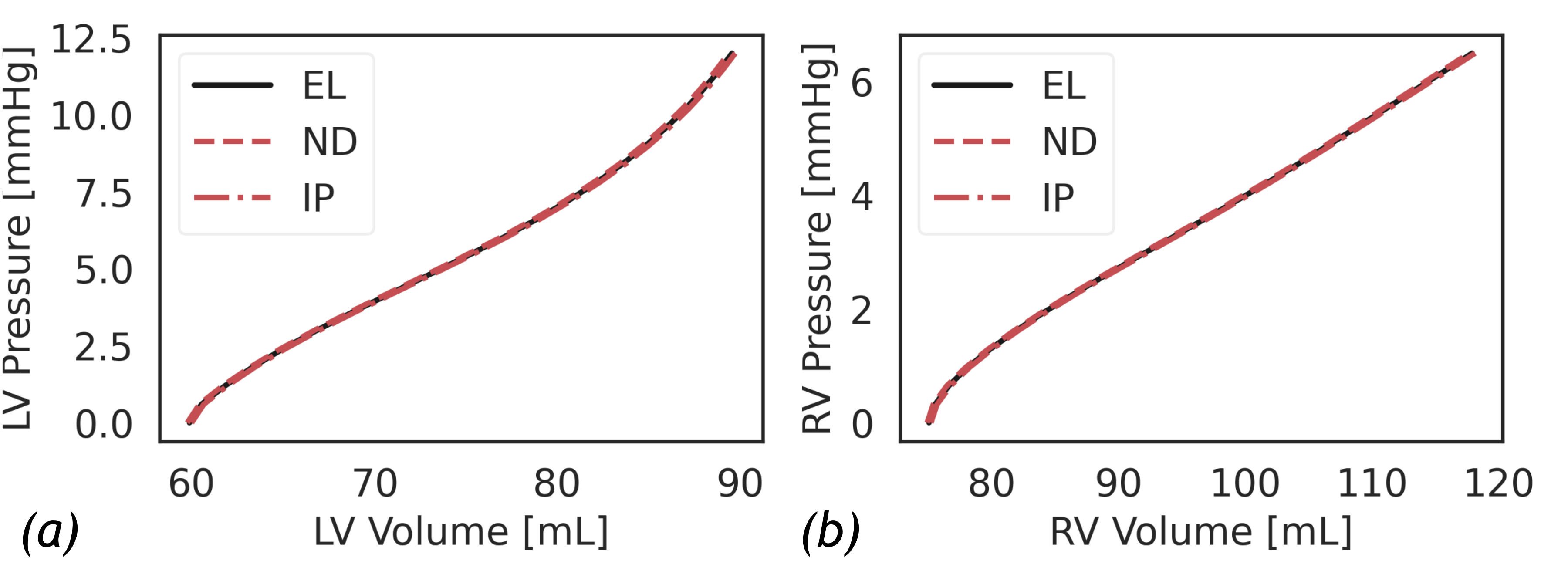}
\caption{Effect of the location for storing the fiber directions on the (a) left and (b) right ventricular pressure-volume (P-V) profiles. No noticeable difference was found by varying the location of fiber directions in our solver. EL: elements; ND: nodes; IP: integration points.}
\label{fig_fiber_locations}
\end{figure}

\subsection{Effects of the location for storing the fiber directions} \label{ssec:fiber_loc}

While performing finite element simulations of fibrous anisotropic media such as the myocardium, one could define the fiber directions at the mesh nodes, as described in Bayer et al.,\cite{bayer2012novel} or at the element centroids. Alternatively, one may define the fiber directions directly at the integration points (i.e., Gauss quadrature points of the element) and use them for evaluating local residue and stiffness tensors via collocation. However, when the fibers are stored at the nodal or element levels, interpolation is typically required to compute fiber directions at the integration points, which are then used to construct the local residue and stiffness tensors. Here, we compared the output of our biventricular mechanics model of the healthy subject by storing the fibers at nodes, element centroids, and integration points (Fig.~\ref{fig_fiber_locations}). Contrary to our expectation that storing fiber directions at the integration points would improve accuracy, we observe that the location of fiber definitions had little effect on the P-V curves (Fig.~\ref{fig_fiber_locations}). This is likely due to the fact that although these fiber directions are stored at different locations, the local residue and stiffness tensors constructed at the element level are projected to the nodes, as all the unknown degrees of freedom are evaluated at the nodes. Redefining the unknown solution vector at the integration points may lower interpolation or projection errors by storing the fibers at the same integration points. However, this would require a massive overhaul of the data structures employed in our code and a substantial increase in the memory requirements, which is beyond this work's scope.

\begin{figure}[ht]
\centering
\includegraphics[width=1\textwidth]{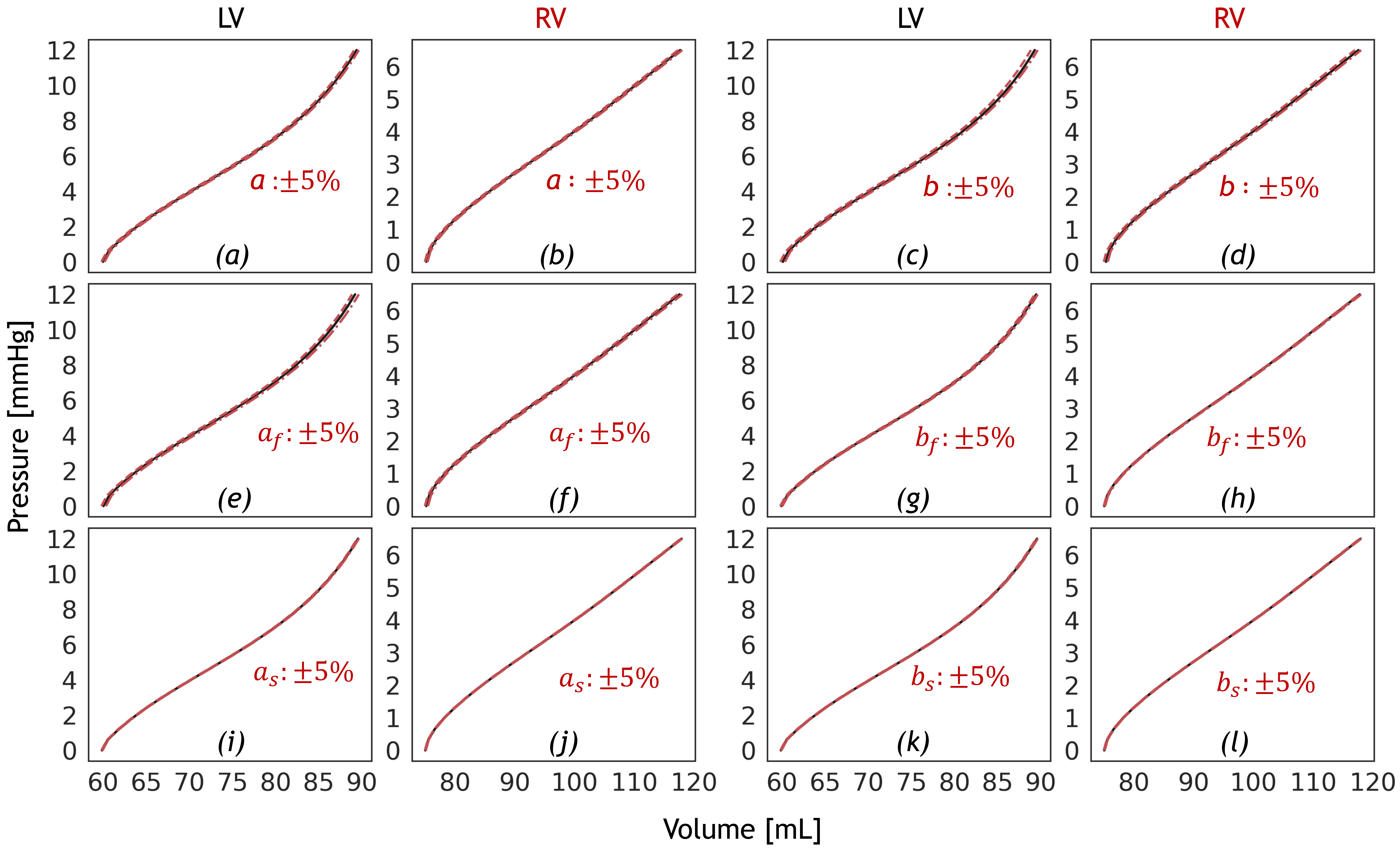}
\caption{Senstivity of the biventricular FEA for a healthy subject to small perturbations in the optimal HO model parameters $\{ a^\mathrm{HO}, b^\mathrm{HO}, a_f^\mathrm{HO}, b_f^\mathrm{HO}, a_s^\mathrm{HO}, b_s^\mathrm{HO} \}$ identified in Table~\ref{table_optim_ifea_params}. The baseline results are shown in a solid black line, while the red dashed and dotted lines indicate a 5\% increase or decrease of the indicated parameter, respectively. Results suggest that small perturbations to the material parameters do not influence the overall P-V profiles. LV: left ventricle; RV: right ventricle.}
\label{fig_sensitivity_study}
\end{figure}

\subsection{Senstivity of the optimal material parameters to small perturbations}\label{ssec:mat_params}

We assess the sensitivity of the biventricular mechanics model to small perturbations ($\pm 5\%$) in each of the estimated optimal material parameters (Table~\ref{table_optim_ifea_params}). Our results suggest that altering each parameter by 5\% increase or decrease does not substantially affect the P-V profiles of both the left and right ventricles (Fig.\ref{fig_sensitivity_study}).

\begin{figure}[ht]
\centering
\includegraphics[width=0.8\textwidth]{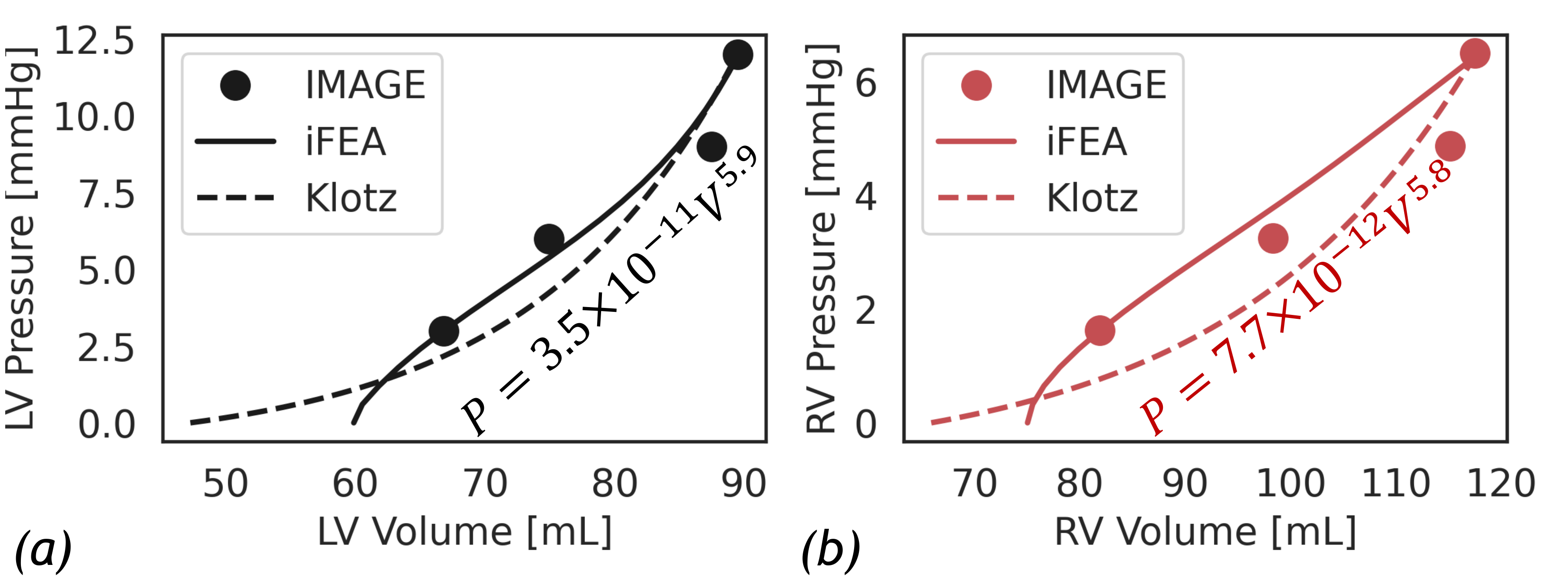}
\caption{Comparison between the P-V profiles derived from the image data, inverse FEA using the HO constitutive model, and the fitted Klotz EDPVR curve for the (a) left and (b) right ventricles of a healthy subject.}
\label{fig_fea_klotz}
\end{figure}

\subsection{Comparison with the Klotz model for EDPVR}\label{ssec:comp_klotz}

The empirical power-law model proposed by Klotz et al.\cite{klotz2007computational} is widely used to approximate the ventricular end-diastolic pressure-volume relation (EDPVR) curve. Originally designed for the left ventricle, we extended the Klotz power-law model to obtain the EDPVR for the left and right ventricles of a healthy subject. This extension allowed us to compare the Klotz-model-based P-V profile against the results obtained through inverse IFEA. The Klotz model of the ventricular P-V curve is given by,

\begin{equation}
   P=\alpha V^{\beta}
\label{eqn_klotz}
\end{equation}

\noindent where $P$ and $V$ are the cavity pressure and volume, respectively. $\alpha$ and $\beta$ are defined as,

\begin{equation}
\begin{split}
   \alpha = &\frac{30}{(V_{30}^{\mathrm{klotz}})^{\beta}}\\      
   \beta = &\frac{\log(P_{\text{ed}}^{\text{img}}/30)}{\log(V_{\text{ed}}^{\text{img}}/V_{\text{30}}^{\text{klotz}})}
\end{split}
\label{eqn_klotz_consts}
\end{equation}

\noindent where, $P_{\text{ed}}^{\text{img}}$ and $V_{\text{ed}}^{\text{img}}$ are the clinically (or experimentally) derived end-diastolic pressure and volume, respectively, and $V_{\text{30}}^{\text{klotz}}$ is defined as

\begin{equation}
V_{\text{30}}^{\text{klotz}} = V_{\text{0}}^{\text{klotz}} + \frac{V_{\text{ed}}^{\text{img}}-V_{\text{0}}^{\text{klotz}}}{(P_{\text{ed}}^{\text{img}}/A_n)^{1/B_n}}
\label{eqn_klotz_v30}
\end{equation}

\noindent where,
\begin{equation}
    V_{\text{0}}^{\text{klotz}}=V_{\text{ed}}^{\text{img}}(0.6-0.006P_{\text{ed}}^{\text{img}})
\label{eqn_klotz_v0}
\end{equation}

\noindent and $A_n=27.78 \text{mmHg}$ and $B_n=2.76$.

For the healthy subject in the current study, the end-diastolic volumes for the left and right ventricles are set as $V_{\text{ed}_{LV}}^{\text{img}} = 89.5\text{mL}$ and $V_{\text{ed}_{RV}}^{\text{img}} = 117.5\text{mL}$, respectively (Table~\ref{table_cavity_vol_img}). Likewise, the end-diastolic cavity pressures are set as $P_{\text{ed}_{LV}}^{\text{img}} = 12 \text{mmHg}$ and $P_{\text{ed}_{RV}}^{\text{img}} = 6.5 \text{mmHg}$, respectively (Section~\ref{ssec:FE_BiV}). These values are substituted in Eqs.~\ref{eqn_klotz}-\ref{eqn_klotz_v0} to obtain the healthy subject’s Klotz-based EDPVR profiles for left and right ventricles (Fig.~\ref{fig_fea_klotz}).

Although both iFEA and Klotz model align reasonably with image data (Fig.~\ref{fig_fea_klotz}), the profile characteristics show subtle differences for the left ventricle (Fig.~\ref{fig_fea_klotz}a) and differ substantially for the right ventricle (Fig.~\ref{fig_fea_klotz}b). These differences are reasonable because the Klotz model assumes a power-law profile and is primarily designed for a left ventricle. Tuning the parameters (Eqs.~(\ref{eqn_klotz_consts}-\ref{eqn_klotz_v0})) may result in a better agreement with the right ventricle. Furthermore, because the Klotz model for pressure-volume relation intercepts the pressure-volume axes only at the origin (Eq.~(\ref{eqn_klotz})), the curve exhibits a slight positive pressure at $V_0^{\mathrm{klotz}}$.

\section{Limitations}\label{sec:limitation}

Here, we acknowledge the current study's limitations to lay a foundation for future work. These limitations generally apply to both biventricular and LA myocardium models, although any limitation specific to a particular model is highlighted.

The current study is focused only on characterizing the passive component of myocardial deformation during the cardiac cycle, excluding the active contraction. However, we note that characterizing passive mechanics is a precursor to characterizing the active component and the entire cardiac cycle. We will pursue this objective in a follow-up investigation using our recently developed modular framework for coupled multiscale cardiac mechanics modeling.\cite{brown2024modular}

Our modeling framework relies on imposing myocardial fiber direction using rule-based methods.\cite{bayer2012novel,piersanti2021modeling} However, as noted earlier in the sensitivity analysis (Section~\ref{ssec:fiber_orient}), the longitudinal fibers' orientation can substantially influence myocardial mechanics. Incorporating realistic fiber distribution through advances in medical imaging, e.g., optical imaging, \textit{in vivo} diffusion tensor MRI,\cite{froeling2014diffusion,khalique2020diffusion,tang2021polarization} and adopting suitable constitutive models,\cite{avazmohammadi2019contemporary} will advance the patient-specific component of the modeling framework.

Due to the unavailability of clinical pressure tracings, we assumed diastolic ventricular cavity pressures based on the literature for healthy individuals and HOCM patients.\cite{klingensmith2008washington, nishimura2017hypertrophic} However, because the diastolic pressures are typically lower in magnitude, we believe the patients’ pressures will be similar to the values utilized in this study. While the utilization of patient-specific pressures might alter the material parameters, integrating such pressures into our modeling framework is straightforward and should not substantially affect the overall methodology. %In an ongoing study on multiscale cardiac mechanics modeling, we use the patient’s cuff pressures to tune the parameters of a multiscale model to replicate the patient’s physiology.

The Robin boundary condition on the basal plane is inadequate to capture ventricular deformation during passive expansion. The ventricular filling during late diastole is dominated by atrial contraction (A-wave) that pushes blood into the ventricular cavity from the atria. As a result, the contracting atria pulls the atrioventricular plane (i.e., basal plane) longitudinally away from the ventricular apex, also evidenced through clinical images.\cite{palit2017passive} However, this pulling effect is not captured by applying a Robin boundary condition alone. On the contrary, due to the imposed stiffness on the basal plane constraining its movement, an increase in the ventricular cavity pressure during atrial filling pushes the ventricular mid-walls outward, like an inflating balloon. Although this basal motion has been captured in prior work by imposing image-based displacements,\cite{palit2017passive} such a model may have limited value for predictive modeling. On the other hand, creating a 4-chamber model with active atrial contraction would allow accurate capturing of atrioventricular interactions and physiological deformation but poses a challenge for inverse modeling owing to the computational expense.\cite{fedele2023comprehensive} A relatively simpler 2-chamber left heart model with a coupled left atrium and left ventricle might be a trade-off between computational cost and replicating the physiological motion of the basal plane.

We assumed a uniform thickness for modeling the left atrial myocardium owing to limitations in the imaging modality to capture the thin tissue. Although this is a usually followed approach,\cite{piersanti2021modeling, hoermann2019automatic} advances in imaging will address this limitation in the future. Further, HOCM tissue is known to be dominated by myocyte disarray and fibrosis.\cite{marian2017hypertrophic, mojumder2023computational} Our model does not account for these effects and will be addressed in future studies. Although the current modeling framework has been applied to multiple patients, the population size must be increased to demonstrate statistical significance in identifying meaningful correlations between material parameters and disease.

%%%%%%%%%%%%%%%%%%%%%%%%%%%%%%%%%%%%%%%%%%%%%%%%%%%%%%%%%%%%%%%%%%%%%%%%%%%
%------------------------    CONCLUSION   -------------------------------
%%%%%%%%%%%%%%%%%%%%%%%%%%%%%%%%%%%%%%%%%%%%%%%%%%%%%%%%%%%%%%%%%%%%%%%%%%%

\section{Conclusion}\label{sec:conclusion}

We propose a novel and efficient inverse finite element analysis (iFEA) framework to characterize the passive cardiac mechanics using time-dependent image data. The iFEA framework is based on a nested optimization scheme in which the outer iterations employ a traditional optimization method to estimate passive material parameters of the myocardium, while the inner iterations employ an augmented Sellier’s method to estimate the stress-free reference configuration. At its core, the framework relies on a recently developed stabilized variational multiscale formulation to simulate the mechanics of biological tissues, which was verified for cardiac mechanics applications. We employ structurally based anisotropic hyperelastic constitutive models and apply physiological boundary conditions to simulate myocardial mechanics. The framework has been tested in models of biventricular myocardium and left atrium in a control subject and three patients with hypertrophic obstructive cardiomyopathy, and demonstrated reasonable agreement in predicting myocardial deformation extracted from the patients’ computed tomographic (CT) images. Our analysis showed that the estimated material parameters are sensitive to myocardial fiber directions and the optimization method, although the P-V profiles and myocardial deformation are robust to the variations in material parameters.

%In conclusion, this study introduces an innovative and efficient optimization scheme aimed at determining the stress-free configuration and optimal material parameters for the patient-specific biventricular model and left atrium model, particularly focusing on the passive process where the deformation is dominated by the pressure provided by blood. Leveraging real image data as the ground truth, the proposed workflow establishes a foundation for developing personalized, multi-physical computational ventricular models. The integration of electrophysiology, ventricular mechanics, and hemodynamics holds promise for advancing precision surgery in the treatment of HOCM. This work sets the stage for further investigations into refining and expanding the computational model to enhance its clinical relevance and utility.

\clearpage

\section*{Acknowledgments}
The authors would like to acknowledge financial support from the American Heart Association’s Second Century Early Faculty Independence Award ($\#$24SCEFIA1260268) in performing this work. The authors would also like to acknowledge computing resources and services received through the Columbia Shared Research Computing Facility and the Ginsburg High-Performance Computing Cluster.

\section*{Conflict of interest statement}
None.

% \bmhead{Supplementary information}

% If your article has accompanying supplementary file/s please state so here.

% \bmhead{Acknowledgments}

% Acknowledgments are not compulsory. Where included they should be brief. Grant or contribution numbers may be acknowledged.



\bibliographystyle{elsarticle-num}
\bibliography{bibliography.bib}





% \end{thebibliography}

%%%%%%%%%%%%%%%%%%%%%%%%%%%%%%%%%%%%%%%%%%%%%%%%%%%%%%%%%%%%%%%%%%%%%%%%%%%
%---------------------------   Appendix   ---------------------------------
%%%%%%%%%%%%%%%%%%%%%%%%%%%%%%%%%%%%%%%%%%%%%%%%%%%%%%%%%%%%%%%%%%%%%%%%%%%

\appendix

\section{Verification of the cardiac mechanics solver}\label{sec:suppl_verify}

While the current stabilized variational multiscale (VMS) formulation for nonlinear structural mechanics has been verified for various canonical solid and fluid mechanics applications,\cite{liu2018unified} here we will verify the solver against cardiac mechanics benchmark problems proposed in Land et al.\cite{land2015verification} In particular, we will employ the VMS formulation to simulate (i) inflation and (ii) inflation and contraction of an idealized left ventricle (i.e., benchmark problems 2 and 3 in Land et al.\cite{land2015verification}).

\begin{figure}[ht]
\centering
\includegraphics[width=0.7\textwidth]{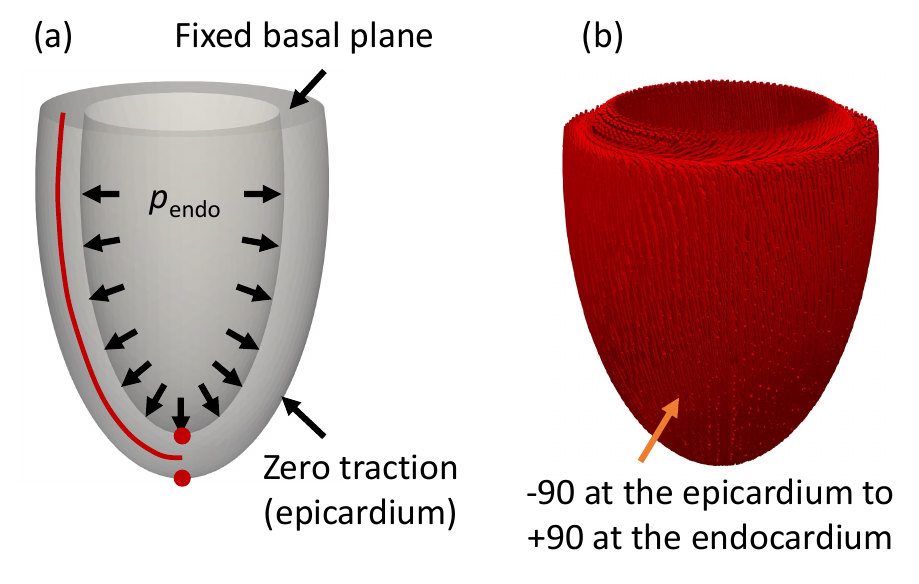}
\caption{A schematic of the truncated ellipsoid used as an idealized left ventricle with (a) boundary conditions and (b) fiber orientations for active contraction.}
\label{fig_suppl_lv}
\end{figure}

In both these cases, the undeformed geometry is described by a truncated ellipsoid (Fig. \ref{fig_suppl_lv}), parameterized as,

\begin{equation}
\mathbf{x} = 
 \left( \begin{array}{c} x\\y\\z \end{array}\right) =
  \left(\begin{array}{c}
   r_s \sin u \cos v    \\ 
   r_s \sin u \sin v  \\  
   r_l \cos u   
   \end{array}\right) 
\label{eq_suppl_ellipsoid}
\end{equation}

\noindent and defined by the volume between:

\begin{itemize}
\setlength{\itemindent}{-1.5em}
\item The \emph{endocardial surface}, $r_s=7\text{mm}, r_l=17\text{mm}, u \in [-\pi, -\arccos \frac{5}{17}], v\in[-\pi,\pi]$, 
\item The \emph{epicardial surface}, $r_s=10\text{mm}, r_l=20\text{mm}, u \in [-\pi, -\arccos \frac{5}{20}], v\in[-\pi,\pi]$, \text{and}
\item The \emph{basal plane}, $z=5\text{mm}$, implicitly defined by the ranges for $u$.
\end{itemize}

A schematic of the model along with imposed boundary conditions is shown in Fig.~\ref{fig_suppl_lv}a. The basal plane is fixed in all directions using a Dirichlet boundary condition, while a traction-free Neumann boundary condition is applied on the epicardial surface. A hemodynamic pressure is applied on the endocardial surface, along the surface normal, that changes in magnitude with deformation (i.e., applied pressure follows deformation). The myocardium is modeled using the transversely isotropic Guccione-McCulloch (GM) constitutive model (Eq.~(\ref{eq_const_gm})).

While Land et al. posed the original problem as statics, here we solve the problem as dynamics with the material density, $\rho_0 = 1\text{g/cm}^3$. We set the time step size to $0.01\text{s}$ and integrate for a total of $200$ time steps using the second-order accurate generalized-$\alpha$ method with spectral radius, $\rho_\infty = 0.5$. The pressure load is incremented linearly from 0 to the peak value for $1\text{s}$, and remains constant at the peak value for the rest of the simulation until the deformation reaches equilibrium. To enforce incompressibility, we use a decoupled form of the strain energy density with the ST91 volumetric strain energy potential, parameterized by the penalty parameter, $\kappa$ (Eq.~(\ref{eq_ST91_Gvol})).

The ellipsoid is discretized with $\sim$22.5K nodes and about 108K linear tetrahedral elements, chosen after a mesh convergence study. The nonlinear system of equations is solved using the Newton-Raphson method, and within each nonlinear iteration, a Krylov space method (GMRES) is used to solve the linear system of equations. A thresholded incomplete LU (ILUT) preconditioner is applied to accelerate the convergence of the GMRES iterative solver, implemented using the Trilinos linear solver library. We set the linear and nonlinear tolerances to $10^{-6}$ for both the benchmark cases.

To assess the performance of the current solver and VMS formulation for cardiac mechanics, we evaluate (a) the deflection of the apex along the apico-basal direction ($z$-axis) at the endocardium and epicardium (marked by thick red circles in Fig.~\ref{fig_suppl_lv}a), and (b) the deformation of the ventricular mid-wall (red curve in Fig.~\ref{fig_suppl_lv}a), and compare against all the participating groups in the original study.\cite{land2015verification}

\subsection{Inflation of an idealized left ventricle}\label{ssec:suppl_benchmark1}

For the passive inflation, the constitutive model parameters are $C=10\text{kPa}, b_f = b_s = b_{fs} = 1$. The base plane($z=5\text{mm}$) is fixed in all directions, while a $10\text{kPa}$ pressure is applied to the endocardium ($p_\text{endo} = 10 \text{kPa}$). The momentum and continuity stabilization parameters, $c_{\tau_m}, c_{\tau_c}$ are set to $0.001$, while the elastic modulus and Poisson's ratio are set to $E = 10\text{kPa}$ and $\nu = 0.5$, respectively. These parameters are used to compute the shear wave speed used in the stabilization coefficients ($\tau_\text{m}, \tau_\text{c}$ in Eq.~\ref{eq_res_stab}). The penalty parameter, $\kappa$, used in the ST91 model is set to $1\text{GPa}$ to penalize incompressible deformations. 

\begin{figure}[ht]
\centering
\includegraphics[width=0.9\textwidth]{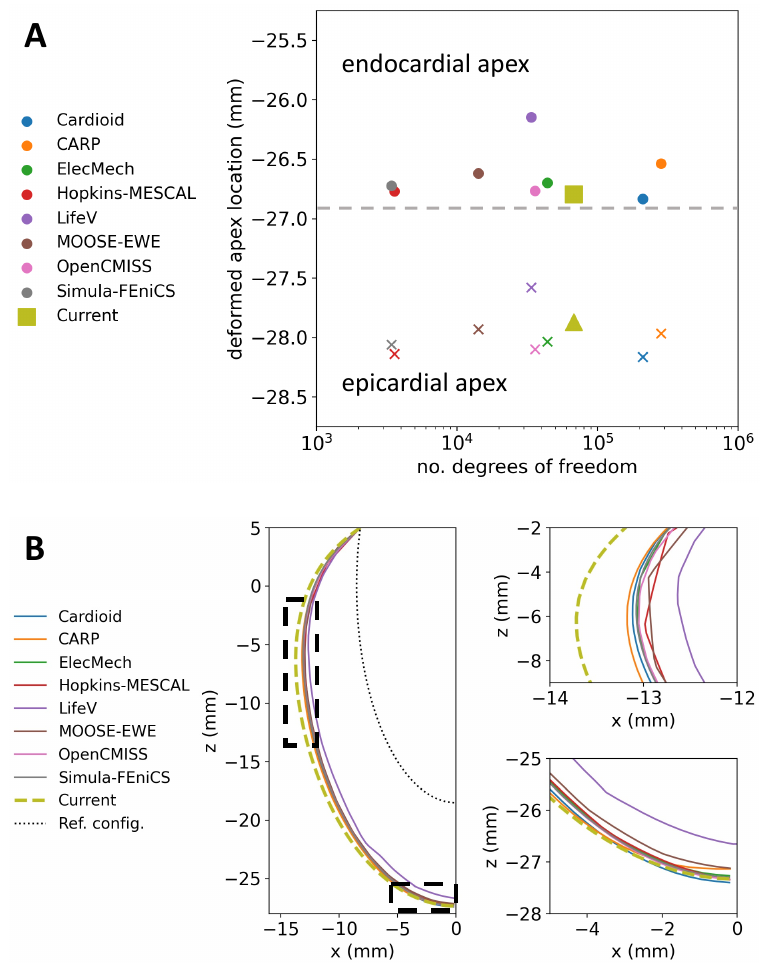}
\caption{Benchmarking LV inflation. Comparison of (A) apex locations and (B) deformation of myocardial mid-line between the current method and other participants in Land et al.\cite{land2015verification}}
\label{fig_suppl_benchmark2}
\end{figure}

We found that the locations of the endocardial and epicardial apexes are well-captured by the current method, agreeing with other participating groups in the original study (Fig. \ref{fig_suppl_benchmark2}A). Likewise, the deformation of the myocardial mid-line from the basal plane to the apex shows a reasonable agreement with other groups (Fig. \ref{fig_suppl_benchmark2}B, left). A careful assessment of the mid-ventricular region shows up to $\sim 0.7\text{mm}$ expansion compared to most other groups (Fig. \ref{fig_suppl_benchmark2}B, upper-right), while the apical region shows very close agreement with a majority of the participants (Fig. \ref{fig_suppl_benchmark2}B, lower-right). We are, therefore, confident that our VMS approach is verified to model passive cardiac mechanics. 

\subsection{Inflation and contraction of an idealized left ventricle}\label{ssec:suppl_benchmark2}

For the combined inflation and active contraction problem, the GM constitutive model parameters are chosen as $C=2\text{kPa}, b_f=8, b_s=2, b_{fs}=4$. The base plane($z=5\text{mm}$) is fixed in all directions, similar to the previous case, while a $15\text{kPa}$ pressure is applied to the endocardium ($p_\text{endo}=15\text{kPa}$). An active stress model is employed to simulate contraction, where the total $2^{nd}$ Piola-Kirchhoff stress is divided into a passive component and an active component, such that,

\begin{equation}
\mathbf{S} = \mathbf{S}_{\text{passive}} + T_a \mathbf{f_0} \otimes \mathbf{f_0}
\label{eqn:suppl_strs}
\end{equation}

\noindent where, $\mathbf{S}_\text{passive}$ is the passive $2^{nd}$ Piola-Kirchhoff stress for the GM constitutive model, and $T_a$ is the active stress acting along the unit fiber direction in the reference configuration. In this problem, $T_a$ is set to $60\text{kPa}$, and its dynamic profile follows pressure load, implying a linear increase for 1s and remains constant after that - both $T_a$ and endocardial pressure ($p_\text{endo}$) are incremented simultaneously.

The fiber directions are computed using a rule-based approach, varying from $-90^\circ$ at the epicardial surface to $+90^\circ$ at the endocardial surface (Fig.~\ref{fig_suppl_lv}b). These are computed using the direction of the derivatives of the ellipsoid parametrization in Eq. \ref{eq_suppl_ellipsoid} as,

\begin{equation}
 \mathbf{f}(u,v) =
   \mathcal{F} \!\left( \frac{ \text{d}\mathbf{x} }{ \text{d}u } \right) \sin \alpha + 
   \mathcal{F} \!\left(\frac{ \text{d}\mathbf{x} }{ \text{d}v } \right)\cos \alpha
\label{eq_suppl_fibers}
\end{equation}

\noindent where, $\mathcal{F}(\mathbf{v}) = \mathbf{v}/\|\mathbf{v}\|$, and the arguments $\frac{\text{d}\mathbf{x}}{\text{d}u}$ and $\frac{\text{d}\mathbf{x}}{\text{d}v}$ are given by,

\begin{equation}
\frac{\text{d}\mathbf{x}}{\text{d}u} = \left(\begin{array}{c} r_s \cos u \cos v \\ r_s \cos u \sin v \\  -r_l \sin u \end{array}\right)  \;,\enspace
\frac{\text{d}\mathbf{x}}{\text{d}v} = \left(\begin{array}{c} -r_s \sin u \sin v \\ r_s \sin u \cos v \\ 0  \end{array}\right)
\end{equation}

\noindent where $r_s = 7 + 3\eta$, $r_l = 17 + 3\eta$, and $\alpha = 90 - 180\eta$ are derived from the transmural distance $\eta \in [0,1]$ varying linearly from $0$ on the endocardium to $1$ on the epicardium.

The momentum and continuity stabilization parameters, $c_{\tau_m}, c_{\tau_c}$ are set to $10^{-5}$, while the elastic modulus and Poisson's ratio are set to $E = 2\text{kPa}$ and $\nu = 0.5$, respectively. These parameters are used to compute the shear wave speed used in the stabilization coefficients ($\tau_\text{m}, \tau_\text{c}$ in Eq.~(\ref{eq_res_stab})). The penalty parameter, $\kappa$, used in the ST91 model is set to $10\text{MPa}$ to penalize incompressible deformations. 

\begin{figure}[ht]
\centering
\includegraphics[width=0.9\textwidth]{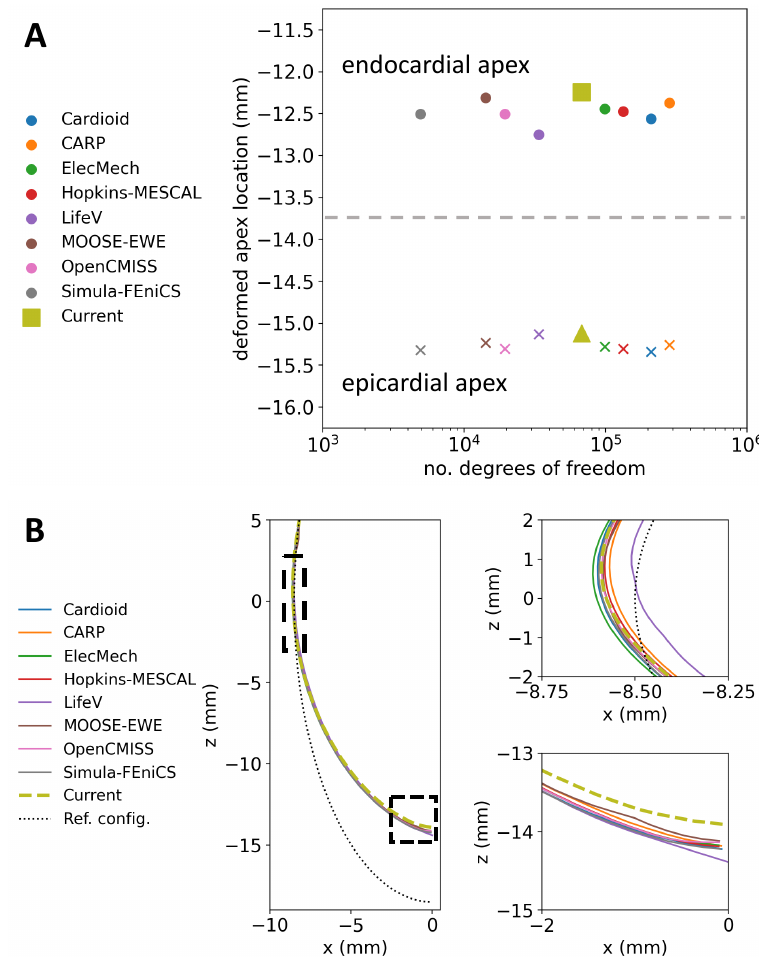}
\caption{Benchmarking LV inflation and contraction. Comparison of (A) apex locations and (B) deformation of myocardial mid-line between the current VMS formulation and other participants in Land et al.\cite{land2015verification}}
\label{fig_suppl_benchmark3}
\end{figure}

We found that the locations of the endocardial and epicardial apexes are adequately captured by the current method compared to other participating groups in the original study (Fig. \ref{fig_suppl_benchmark3}A). Likewise, the deformation of the myocardial mid-line from the basal plane to the apex shows excellent agreement with other groups (Fig. \ref{fig_suppl_benchmark3}B, left). The ventricular deformation at the mid-ventricular region overlaps with most participants (Fig. \ref{fig_suppl_benchmark3}B, upper-right), while the apical region underpredicts ventricular contraction by $0.2-0.3\text{mm}$ compared to other groups (Fig. \ref{fig_suppl_benchmark3}B, lower-right). This exercise demonstrates that our current formulation for solving cardiac mechanics is verified to capture both passive inflation and active contraction.

\clearpage

\end{document}